\documentstyle{article}
\newcommand\mysection{\setcounter{equation}{0}\section}
\renewcommand{\theequation}{\thesection.\arabic{equation}}
\newcounter{hran} \renewcommand{\thehran}{\thesection.\arabic{hran}}

\def\bmini{\setcounter{hran}{\value{equation}}
  \refstepcounter{hran}\setcounter{equation}{0}
  \renewcommand{\theequation}{\thehran\alph{equation}}\begin{eqnarray}}

\def\bminiG#1{\setcounter{hran}{\value{equation}}
\refstepcounter{hran}\setcounter{equation}{-1}
\renewcommand{\theequation}{\thehran\alph{equation}}
\refstepcounter{equation}\label{#1}\begin{eqnarray}}

\def\emini{\end{eqnarray}\relax\setcounter{equation
}{\value{hran}}\renewcommand{\theequation}{\thesection.\arabic{equation}}}

\def\beq{\begin{equation}}
\def\eeq{\end{equation}}
\newcommand{\be}{\begin{equation}}
\newcommand{\ee}{\end{equation}}
\def\bea{\begin{eqnarray}}
\def\eea{\end{eqnarray}}

\newcommand{\barre}[1]{%
	\setbox1=\hbox{$#1$} \dimen2=\ht1 \dimen3=\dp1 \dimen4=\wd1
	\setbox2=\hbox{\sl /}
	\dimen1=\wd1 \advance\dimen1 by -\wd2 \divide\dimen1 by 2
	\advance\dimen1 by \wd2 \advance\dimen1 by 0.4pt
	\setbox3=\hbox to \wd1{\hss \box1 \kern -\dimen1 \box2\hss}
	\ht3=\dimen2 \dp3=\dimen3 \wd3=\dimen4
	\box3
	}

\newcommand{\vev}[1]{%
	\langle #1 \rangle
	}
\newcommand{\svev}[1]{%
	|\langle #1 \rangle|^2
	}

\def\1{{\rm 1 \kern -.10cm I \kern .14cm}} \def\R{{\rm R \kern -.28cm I
\kern .19cm}}

\begin{document}

\begin{titlepage}
\begin{flushright}    UFIFT-HEP-98-06 \\ 
\end{flushright}
\vskip 1cm
\centerline{\LARGE{\bf {Predictions from an Anomalous $U(1)$ }}}
\vskip .5cm 
\centerline{\LARGE{\bf {Model of Yukawa Hierarchies}}}
\vskip 1.5cm
\centerline{\bf Nikolaos Irges${}^{\,}$\footnote{Supported in part by the
United States Department of Energy under grant DE-FG02-97ER41029.}, St\'ephane 
Lavignac${}^{\,}$\footnote{Attach\'e Temporaire d'Enseignement et de 
Recherche, Universit\'e Paris VII, Paris, France.}${}^{,}
$\footnote{Permanent address: Laboratoire de Physique Th\'eorique et Hautes
Energies, Universit\'e Paris-Sud, B\^at. 210, F-91405 Orsay Cedex, France.}, 
and Pierre Ramond${}^{\, 1}$}   
\vskip .5cm
\centerline{\em  Institute for Fundamental Theory,}
\centerline{\em Department of Physics, University of Florida}
\centerline{\em Gainesville FL 32611, USA}
\vskip 1.5cm

\centerline{\bf {Abstract}}
\vskip .5cm
We present a supersymmetric standard model with three gauged Abelian 
symmetries, of a type commonly found in superstrings.  
One is anomalous, the other two are $E_6$ family symmetries. It has a  
vacuum in which only these symmetries are broken by stringy effects.  It reproduces
all observed quark and charged lepton Yukawa hierarchies, and the value 
of the Weinberg angle. It  predicts three
massive neutrinos, with  mixing that can explain both the small angle MSW 
effect, and the atmospheric neutrino anomaly.  The Cabibbo angle is 
expressed in terms of the gauge couplings at unification. It conserves 
R-parity, and proton decay is close to experimental bounds. 

\vfill
\begin{flushleft}
February 1998 \\
\end{flushleft}
\end{titlepage}


\mysection{Introduction}

Over the last few years, there has been growing interest in relating 
generic features of superstring models to low energy phenomenology. 
Prominent among these, are models which contain an anomalous $U(1)$ with 
anomalies cancelled by the Green-Schwarz mechanism~\cite{GS}, and in 
which the dilaton 
gets a vacuum value, generating a Fayet-Iliopoulos term that triggers the 
breaking~\cite{DSW} of at least the anomalous gauged symmetry at a  large 
computable scale. 

Through the anomalous $U(1)$, the Weinberg angle at the cut-off is 
related to anomaly coefficients~\cite{Ib}. This allows for  possible relations 
between fundamental string quantities (in the ultraviolet) and 
experimental parameters (in the infrared).  A simple model~\cite{IR} with one 
family-dependent anomalous $U(1)$ beyond the standard model was the first 
to exploit these features to produce Yukawa hierarchies and fix the 
Weinberg angle. It was soon 
realized that  some features could be abstracted from the 
presence of the anomalous $U(1)$: expressing the ratio of down-like quarks to 
charged lepton masses in terms of the Weinberg angle~\cite{BR,Nir,JS}, the suppression of 
the bottom to the top quark masses~\cite{PMR}, relating the uniqueness of the vacuum 
to Yukawa hierarchies and the 
presence of MSSM invariants in the superpotential, and finally  
relating the see-saw mechanism~\cite{SEESAW} to R-parity 
conservation~\cite{BLIR}. 

Recently, many of these ideas were incorporated in a model~\cite{EIR} with 
one anomalous and two  non-anomalous $U(1)$ symmetries spontaneously broken 
by stringy effects. It contained only the three standard model chiral 
families, three right-handed neutrinos, and the fields necessary to break 
the extra phase symmetries. It reproduced all quark and 
charged lepton hierarchies, and the Weinberg angle, but failed in some 
other aspects: the proton decayed faster than observed, and the three 
light neutrinos had an inverse mass hierarchy, which could not account 
for the  solar neutrino deficit.

In this paper, we propose an alteration of this model, in which there 
are  two non-anomalous family symmetries contained within $E_6$, and one 
anomalous family-independent symmetry. All three are spontaneously broken 
by the dilaton-generated FI term. To cancel anomalies, it contains 
vector-like matter with standard model charges, and hidden sector fields 
and interactions, some of the  many features encountered in superstring 
models. It is expressed as an effective low-energy supersymmetric theory 
with a cut-off scale $M$. It  has some distinctive features, such as:

\begin{itemize}
\item All quark and charged lepton Yukawa hierarchies, and mixing, 
including the bottom to top Yukawa suppression. 

\item The value of the Weinberg angle at unification.   

\item Three massive neutrinos with mixings that give the small-angle 
MSW effect for the solar neutrino deficit, and the large angle mixing 
necessary for the atmospheric neutrino effect.

\item Natural R-parity conservation.

\item Proton decay into $K^0+\mu^+$ near the experimental limit.

\item A hidden sector that contains  strong gauge interactions. 

\end{itemize}
It is heavily constrained by the requirement that the vacuum, in which the 
three $U(1)$'s are broken by stringy effects, be free of  flat directions 
associated with the MSSM invariants. Our model's vacuum is demonstrably 
free of the flat direction associated with each invariant.  

The theoretical consistency of the model is tested by the many ways in which its 
cut-off is ``measured". First, the renormalization group evolution of the 
standard model gauge couplings yields their unification scale. Its value 
depends on the number of standard-model vector-like matter at 
intermediate masses; in our model we find $M_U\sim 3\times 10^{16}$ GeV. 
Secondly, assuming that all couplings in the superpotential are of order 
one, it is measured by fitting the neutrino mass scale. A fit to both the 
small angle MSW and the atmospheric neutrino deficit yields 
$10^{16}<M<4\times 10^{17}$ GeV. A fit only to the MSW effect yields 
a larger value, $M\sim 10^{18}$ GeV. Thirdly,  the lack of 
experimental evidence for proton decay sets a lower bound for $M$ 
consistent with these estimates.

In all above estimates, we have used the Cabibbo angle as expansion 
parameter. However, the Green-Schwarz relation 
 yields a natural expansion parameter in terms of the gauge coupling at 
unification. In our model, we find it to be $\lambda\sim .28$, clearly of
the same order of magnitude but larger than the Cabibbo angle, but 
this value depends on the standard-model vector-like matter, about which 
we have no direct experimental information. Thus we 
have used the experimental value of the Cabibbo angle in all estimates of 
the suppression factors.  

Furthermore, our determination~\cite{Ib} of the Weinberg angle assumes 
that the cut-off is  close to the unification scale. Thus  
  theoretical and numerical considerations imply that if our theory is 
to be derived from a theory in higher dimensions, its ``string" cut-off 
must be near the 
unification scale.

To complete our model, we need to include mechanisms that break both 
supersymmetry and electroweak symmetries. The  hidden sector contains a  
gauge theory with strong coupling,  capable of breaking supersymmetry 
through the Bin\'etruy-Dudas mechanism~\cite{BD}. Unfortunately, it 
cannot be the main agent of supersymmetry breaking. The reason is that 
squarks get soft masses through the D-terms of the gauge symmetries, and 
while the D-term of the family-independent anomalous $U(1)$ yields equal 
squark masses, the D-terms of the other two $U(1)$'s give generically 
flavor-dependent contributions$\footnote{E. Dudas, private 
communication.}$. Since our model  does not align~\cite{LNS} the quark 
and squark mass matrices sufficiently to account for the flavor-changing 
constraints, we are left  with the usual flavor problem associated with 
supersymmetry-breaking. Also, this mechanism   does not generate large 
gaugino masses. We note that in some free-fermion superstring 
models~\cite{faraggi},  the flavor-dependent D-terms can vanish

In the following, we present the details of the model. Section 2 details 
the gauge sector, followed in Section 3, by a discussion of the gauge 
anomalies, and their cancellations. This is followed in Section 4 by a 
discussion of the general features of its vacuum. The phenomenology of 
quark and lepton masses is presented in Section 5, followed in Section 6 
by a thorough discussion of the neutrino phenomenology of our model.  In 
Section 7, we analyze the consequences of the matter with vector-like 
standard model charges. The discussion of the matter content concludes in 
Section 8 with the hidden sector needed to cancel anomalies. We then 
describe in Section 9 how R-parity conservation arises in our model, 
followed in Section 10 by the analysis of proton decay interactions. 
Finally we close with a detailed analysis of the vacuum flat directions 
associated with the invariants of the model.


\mysection{The Gauge Sector}

In the visible sector, the gauge structure of our model is that of the 
standard model, augmented by three Abelian symmetries:

\beq SU(3)\times SU(2)\times U(1)_Y\times U(1)_X\times U(1)_{Y^{(1)}}\times  
U(1)_{Y^{(2)}}\ .\eeq  
One of the extra symmetries, which we call $X$, is anomalous in the sense of 
Green-Schwarz; Its charges are assumed to be family-independent. The other two 
symmetries, $Y^{(1)}$ and $Y^{(2)}$, are not anomalous, but have specific 
dependence on the three chiral families, designed to reproduce the Yukawa 
hierarchies. This theory is inspired 
by models generated from the superstring $E_8\times E_8$ heterotic 
theory, and its chiral matter lies in broken-up representations 
of $E_6$, resulting in the cancellation of many anomalies. This also 
implies the presence of  both matter that is vector-like with respect to 
standard model charges, and  right-handed neutrinos, which trigger 
neutrino masses through the seesaw mechanism~\cite{SEESAW}.   

The three symmetries, $X$, $Y^{(1,2)}$ are spontaneously broken at a high 
scale by the Fayet-Iliopoulos term generated by the dilaton vacuum. This 
DSW vacuum~\cite{DSW} is required to preserve both supersymmetry and the 
standard model symmetries. Below its scale, our model displays only the 
standard model gauge symmetries.

To set our notation, and explain our charge assignments, let us recall 
some basic $E_6$~\cite{ESIX}. It contains two Abelian symmetries outside of
the standard model: 
The first $U(1)$, which we call $V'$, appears in the embedding 

\be E_6\subset ~SO(10)\times U(1)_{V'}\ee
with

\be{\bf 27}={\bf 16}_1+{\bf 10}_{-2}+{\bf 1}_4\ ,\ee
where the $U(1)$ value appears as a subscript. The second $U(1)$, called 
$V$, appears in

\be SO(10)\subset SU(5)\times U(1)_V\ ,\ee
corresponding to

\be{\bf 16}={\bf \overline 5}_{-3}+{\bf 10}_1+{\bf 1}_5\ ;
\qquad {\bf 10}={\bf \overline
5}_{2}+{\bf 5}_{-2}\ .\ee
The familiar hypercharge, $Y$, appears in

\be SU(5)\subset SU(2)\times SU(3)\times U(1)_Y\ ,\ee
with the representation content
\bea
{\bf \overline 5}&
=&({\bf 2},{\bf 1}^c)_{-1}+({\bf 1},{\bf\overline 3}^c)_{2/3}\ ,\\
{\bf 10}&=&({\bf 1},{\bf 1}^c)_2+({\bf 2},{\bf 3}^c)_{1/3}+
({\bf 1},{\bf\overline 3}^c)_{-4/3}\ .
\eea
The two $U(1)$s in $SO(10)$, can also be identified with baryon number 
minus lepton number and right-handed isospin as

\be B-L={1\over 5}(2Y+V)\ ;\qquad I_{3R}={1\over 10}(3Y-V)\ .\ee
The first combination is $B-L$ only on 
the standard model chiral families in the $\bf 16$;  on the 
vector-like matter in the $\bf 10$ of $SO(10)$ it cannot be interpreted 
as their baryon number minus their lepton number. 

We postulate the two non-anomalous symmetries to be  

\beq Y^{(1)}={1\over 5}(2Y+V)
\left( \begin{array}{ccc}
    2 & 0 & 0 \\
    0 & -1 & 0 \\
    0 & 0 & -1   \end{array}  \right)\eeq

\beq Y^{(2)}={1\over 4}(V+3V')\left( \begin{array}{ccc}
    1 & 0 & 0 \\
    0 & 0 & 0 \\
    0 & 0 & -1   \end{array}\right)\ ,\eeq
The family matrices run over the three chiral families, so that $Y^{(1,2)}$ 
are family-traceless. 

We further assume that the $X$ charges on the three chiral families in the
$\bf 27$ are of the form

\be X=(\alpha+\beta V+\gamma V')\left( \begin{array}{ccc}
    1 & 0 & 0 \\
    0 & 1 & 0 \\
    0 & 0 & 1   \end{array}\right)\ ,\eeq
where $\alpha,~\beta,~\gamma$ are as-of-yet undetermined parameters.
Since ${\rm Tr}(YY^{(i)})={\rm Tr}(YX)=0$, there is no appreciable 
kinetic mixing between the hypercharge and the three gauged symmetries. 

\noindent The matter content of this model is the smallest that reproduces the 
observed quark and charged lepton hierarchy, cancels the anomalies 
associated with the extra gauge symmetries, and produces a unique vacuum 
structure:
\begin{itemize}
\item Three chiral families each with the quantum numbers of 
a $\bf 27$ of $E_6$. This means  three chiral families of the standard 
model, ${\bf Q}_i$, $\overline{\bf u}_i$, $\overline{\bf d}_i$, $L_i$, 
and $\overline e_i$, together with three right-handed neutrinos
$\overline N_i$, 
three vector-like pairs denoted by $E_i$ + $\overline{\bf D}_i$ 
and $\overline E_i$ + ${\bf D}_i$, with the quantum numbers of the
$\overline{\bf 5}$ + $\bf 5$ of $SU(5)$. Our model does not contain the 
singlets $S$ that make up the rest of the $\bf 27$. With our charges, 
they are not required  by anomaly cancellation,  and their presence would 
create unwanted flat directions in the vacuum.
\item One standard-model vector-like pair of  Higgs 
weak doublets.
\item Chiral fields that are needed to break 
the three extra $U(1)$ symmetries in the DSW vacuum. We denote these 
fields by $\theta_a$. In our minimal model with three symmetries that 
break through the FI term, we just take $a=1,2,3$. The 
$\theta$ sector  is necessarily anomalous.
 \item Hidden sector gauge interactions and their matter, together with 
singlet fields, needed to cancel the remaining anomalies.
\end{itemize}


\mysection{Anomalies}

In a four-dimensional theory, the Green-Schwarz anomaly compensation 
mechanism occurs through a dimension-five term that couples an axion to 
all the gauge fields. As a result, any anomaly linear in the 
$X$-symmetry must satisfy the Green-Schwarz relations

\be 
(XG_iG_j)=\delta_{ij}C_i\ ,\ee
where $G_i$ is any gauge current. The anomalous symmetry must have a 
mixed gravitational anomaly, so that 

\be
(XTT)=C_{\rm grav}\ne 0\ ,\ee
where $T$ is the energy-momentum tensor. In addition, the anomalies compensated 
by the Green-Schwarz mechanism satisfy the universality conditions

\be
{C_i\over k_i}={C_{\rm grav}\over 12}\ {\rm 
~~~for~all~}~i\ .\ee
A similar relation holds for $C_X\equiv(XXX)$, the self-anomaly
coefficient of the $X$ 
symmetry.  These result in important numerical constraint, 
which can be used to restrict the matter content of the model.

All other anomalies must vanish:
\be
(G_iG_jG_k)=(XXG_i)=0\ .\ee
In terms of the standard model, the vanishing anomalies are therefore of 
the following types: 
\begin{itemize}
 \item  The first involve only standard-model gauge groups $G_{\rm SM}$, 
with coefficients $(G_{\rm SM}G_{\rm SM}G_{\rm SM})$, which cancel for 
each chiral family and for vector-like matter. Also the hypercharge mixed 
gravitational anomaly $(YTT)$ vanishes.
\item The second type is where the new symmetries 
appear linearly, of the type $(Y^{(i)}G_{\rm SM}G_{\rm SM})$. The choice 
of family-traceless $Y^{(i)}$ insures their vanishing  over the 
three families of  fermions with standard-model. Hence they must vanish 
on the Higgs fields: with  $G_{\rm SM}=SU(2)$, it implies the Higgs pair is 
vector-like  with respect to the $Y^{(i)}$. It follows that the mixed 
gravitational anomalies $(Y^{(i)}TT)$ are zero over the fields with 
standard model quantum numbers. They must therefore vanish as well over 
all other fermions in the theory.
 
\item The 
third type involve the new symmetries quadratically, of the form $(G_{\rm 
SM}Y^{(i)}Y^{(j)})$. These vanish automatically except for those 
 of the form $(YY^{(i)}Y^{(j)})$. Two types of 
fermions contribute: the three chiral families and 
standard-model vector-like pairs 

\be 0=(YY^{(i)}Y^{(j)})=(YY^{(i)}Y^{(j)})_{\rm 
chiral}+(YY^{(i)}Y^{(j)})_{\rm real}\ .\ee 
By choosing $Y^{(1,2)}$  in $E_6$, 
overall cancellation is assured, but the
vector-like matter is necessary to 
cancel one of the anomaly coefficient, since we have

\be
(YY^{(1)}Y^{(2)})_{\rm chiral}=-(YY^{(1)}Y^{(2)})_{\rm real}=12\ .\ee
\item The fourth type are the anomalies of the new symmetries of the form 
$(Y^{(i)}Y^{(j)}Y^{(k)})$. 
Since standard-model  singlet fermions can contribute, it is not clear 
without a full theory, to determine how the cancellations come about. 
We know that over the fermions in an $E_6$ representation, 
they vanish, but, as  we shall see, the $\theta$ sector is necessarily 
anomalous. In the following we will present a scenario for these 
cancellations, but it is the least motivated sector of the theory since 
it involves the addition of fields whose sole purpose 
is to cancel anomalies.

\item The remaining vanishing anomalies involve the anomalous charge $X$. 
\begin{itemize} \item Since both $X$ and $Y$ are family-independent, and 
$Y^{(i)}$ are family-traceless, the vanishing of the 
$(XYY^{(1,2)})$ coefficients   over the three families is assured, so 
they must vanish over the Higgs 
pair. This means that $X$ is vector-like on the Higgs pair. 
It follows that the standard-model invariant $H_uH_d$ (the $\mu$ 
term) has zero $X$ and $Y^{(i)}$ charges; it can appear by itself 
in the superpotential, but we are dealing with a 
string theory, where mass terms do not appear in the superpotential: it 
can appear only in the K\"ahler potential. This results, after 
supersymmetry-breaking in 
an induced $\mu$-term, of weak strength, as suggested by Giudice and 
Masiero~\cite{GM}. 

Since the Higgs do not contribute to anomaly coefficients, 
 we can compute  the standard model anomaly coefficients. We find

\be
 C_{\rm color}=18\alpha\ ;\ \
 C_{\rm weak}=18\alpha\ ;\ \
 C_Y=30\alpha\ .\ee

Applying these to the Green-Schwarz relations we find the Kac-Moody
levels for the color and weak groups to be the same

\be
k_{\rm color}=k_{\rm weak}\ ,\ee
and through the Ib\'a\~nez  relation~\cite{Ib}, the value of the
Weinberg angle at the cut-off

\be
\tan^2\theta_w={C_{\rm Y}\over C_{\rm weak}}={5\over 3}\ ,\ee
not surprisingly the same value as in $SU(5)$ theories.

\item The  coefficients $(XY^{(1)}Y^{(2)})$. Since 
standard-model singlets can contribute, we expect its cancellation to 
come about through a combination of hidden sector and singlet fields. Its 
contribution over the chiral fermions (including the right-handed 
neutrinos) is found to be 

\be
(XY^{(1)}Y^{(2)})_{\rm chiral~+~real}=18\alpha\ .\ee
\item The coefficient $(XXY)$. With our  choice for $X$, it is zero. 
\item The coefficients $(XXY^{(i)})$ vanish over the three families of 
fermions with standard-model charges, but contributions are expected from 
other sectors of the theory.
\end{itemize}
\end{itemize}
The vanishing of these anomaly coefficients is highly non-trivial, and it 
was the main motivator for our (seemingly arbitrary) choices of $X$, and 
$Y^{(i)}$.


\mysection{The DSW vacuum}
\label{section:FD}

The $X$, $Y^{(1)}$ and $Y^{(2)}$ Abelian symmetries are spontaneously broken 
below the cut-off. Phenomenological considerations require that neither 
supersymmetry nor any of the standard model symmetries be broken at that 
scale. This puts severe restrictions on the form of the superpotential 
and the  matter fields \cite{BLIR}. 

Since three symmetries are to be broken, we assume that three fields, 
$\theta_a$, acquire a vacuum value as a result of the FI term. They are 
singlets under the standard model symmetries, but not under $X$ and 
$Y^{(1,2)}$. If more 
fields than broken symmetries assume non-zero values in the DSW vacuum, 
we would have undetermined flat directions and hierarchies,  and  
Nambu-Goldstone bosons associated with the extra symmetries.

We express their charges in terms of a $3\times 3$ matrix $\bf A$, whose 
rows are the $X$, $Y^{(1)}$ and $Y^{(2)}$ charges of the three
$\theta$ fields, respectively.

Assuming the existence of a supersymmetric  vacuum where only the $\theta$
fields have vacuum values, implies from the vanishing of the three 
$D$ terms 

\be 
{\bf 
A}\left(\begin{array}{c}\vert\theta_1\vert^2\\\vert\theta_2\vert^2\\
\vert\theta_3\vert^2
\end{array}  \right)=\left(\begin{array}{c}\xi^2\\ 0\\ 
0\end{array}\right)\ .\ee
It follows that the matrix $\bf A$ must not only have an inverse but that 
the entries in the first row of its inverse be positive. 

Since ${\bf A}$ is invertible, its rows consist of three linearly independent 
(but not orthogonal) basis vectors, ${\bf v}_1$, ${\bf v}_2$ and ${\bf v}_3$, 
whose components are the $X,~Y^{(1)},~Y^{(2)}$ 
charges of the $\theta$ fields. The charges of any standard model invariant
$S$ (or any standard model singlet $\chi$)
form a vector which can be expressed in that basis:

\be
{\bf w}=-(n_1{\bf v}_1+n_2{\bf v}_2+n_3{\bf v}_3)
\label{eq:powers} \ee
If all $n_\alpha$, $\alpha=1,2,3$ are positive integers, then $S\, \theta^{n_1}_1 
\theta^{n_2}_2 \theta^{n_3}_3$ is a holomorphic invariant and can be present in 
the superpotential. It is quite remarkable that the invertibility of ${\bf A}$, 
which ensures the existence of the DSW vacuum, is the same condition required for 
invariants of the form $S\, \theta^{n_1}_1 \theta^{n_2}_2 \theta^{n_3}_3$ to exist.  
Those invariants are precisely the ones needed to generate mass hierarchies in the 
DSW vacuum, with $S$ being Yukawa invariants. If all $n_{\alpha}$ are positive, but 
some of them are fractional, the invariant appears at higher order: $\left( S\, 
\theta^{n_1}_1 \theta^{n_2}_2 \theta^{n_3}_3 \right)^m$. Finally, if some $n_{\alpha}$ 
is negative, one cannot form any holomorphic invariant out of $S$ and the $\theta$ fields.

We have found no fundamental principle that fixes the charges of the 
$\theta$ fields. However, by 
requiring that they all get the same vacuum value and reproduce the quark 
hierarchies, we arrive at the simple assignment

\be
{\bf A}= \left( \begin{array}{ccc}
1&0&0\\ 0&-1&1\\ 1&-1&0
\end{array}  \right)\ .
\ee
Forming its inverse,  
\be
{\bf A}^{-1}= \left( \begin{array}{ccc}
1&0&0\\ 1&0&-1\\ 1&1&-1
\end{array}  \right)\ ,\ee
we see that all  three $\theta$ fields have the same vacuum expectation value
\be
\vert<\theta_1>\vert=\vert<\theta_2>\vert=\vert<\theta_3>\vert=
\xi \ .\ee

The presence of other fields that do not get values in the DSW 
vacuum severely restricts the form of the superpotential. In particular,
when the extra 
fields are right-handed neutrinos, the uniqueness of the DSW vacuum is 
attained only after  adding to the superpotential terms of the form $\overline 
N^p{\cal P}(\theta)$, where $p$ is an integer $\ge 2$, and $\cal P$ is a 
holomorphic polynomial in the $\theta$ fields. If  
$p= 1$,  its F-term  breaks supersymmetry at the DSW 
scale.

The case $p=2$ is more desirable since it translates into a Majorana mass 
for the right-handed neutrino, while the cases $p\ge 3$ leave the 
$\overline N$ massless in the DSW vacuum.   To single out  
$p=2$ we simply  choose the $X$ charge 
of the $\overline N_i$ to be a negative half-odd integer.  Since 
right-handed neutrinos couple to the standard model invariants
$L_iH_u$, it implies that $X_{L_iH_u}$ is also a half-odd
integer. 

The same analysis can be applied to the invariants of the MSSM.
Since they must be present  in the superpotential 
to give quarks and leptons their masses,
their $X$-charges must be negative integers. Remarkably, 
these are the very same conditions  necessary to avoid  flat
directions along which these invariants do not vanish: with negative
charge, these invariants cannot be the only  contributors to  $D_X$ in
the DSW vacuum. The presence of a holomorphic invariant, linear in
the MSSM invariant multiplied by a polynomial in the $\theta$ fields,
is necessary to
avoid a flat direction where both the invariant and the $\theta$
fields would get DSW vacuum values. 
The full analysis of the DSW vacuum in our model is rather involved,
but it is greatly simplified by using the general methods introduced
by two of us~\cite{IL}. We postpone the discussion of the uniqueness of
the vacuum until the end of this paper.

Finally, we note a curious connection between the DSW vacuum and the
anomalies carried 
by the $\theta$ fields. Assume that the $\theta$ sector does not contribute to 
the mixed 
gravitational anomalies

\be
(Y^{(i)}TT)_\theta=0\ .\ee
This means that the charges $Y^{(i)}$ are traceless over the $\theta$ sector. 
They are therefore generators of the global $SU(3)$ under which the three 
$\theta$ fields form the  $\bf 3$ representation. However, $SU(3)$ is 
anomalous, and it contains only one non-anomalous $U(1)$ that resides in 
its $SU(2)$ subgroup. Thus to avoid anomalies, the two charges $Y^{(1,2)}$ 
need to be aligned over the 
$\theta$ fields, but this would imply $\det{\bf A}=0$, in 
contradiction with the necessary condition for the DSW vacuum.
It follows that the vacuum structure {\em requires} the $\theta$ sector to be 
anomalous. Indeed we find that, over the $\theta$ fields,

\be
(Y^{(1)}_{}Y^{(1)}_{}Y^{(2)}_{})_\theta 
=(Y^{(1)}_{}Y^{(2)}_{}Y^{(2)}_{})_\theta=-1\ .
\ee
In a later section we discuss how these anomalies might be compensated.


\mysection{Quark and Charged Lepton Masses}

To account for the top quark mass, we assume that the superpotential
contains the invariant

\be
{\bf Q}_3{\overline{\bf u}_3}H_u\ .\ee
Since $X$ is family-independent, it follows that the standard-model 
invariant operators ${\bf Q}^{}_i\bar{\bf u}^{}_jH^{}_u$, where $i,j$ are 
family indices, have zero $X$-charge. Together with the anomaly 
conditions, this fixes the Higgs charges 

\be X_{H_u}=-X_{H_d}=-X_{\bf Q}-X_{\overline{\bf u}}\ ,\ee
and
\be Y^{(1)}_{H_u}=-Y^{(1)}_{H_d}=0\ ,\qquad 
Y^{(2)}_{H_u}=-Y^{(2)}_{H_d}=2\ .\ee

\be
X({\bf Q}^{}_i\bar{\bf u}^{}_jH^{}_u)\equiv X^{[u]}=0\ .\ee
The superpotential contains terms of higher dimensions. In the 
charge $2/3$ sector, they are

\be {\bf Q}^{}_i\bar{\bf u}^{}_jH^{}_u
{\bigl ( {\theta_1 \over M} \bigr )}^{n^{(1)}_{ij}}
{\bigl ( {\theta_2 \over M} \bigr )}^{n^{(2)}_{ij}}
{\bigl ( {\theta_3 \over M} \bigr )}^{n^{(3)}_{ij}}
\ ,\label{eq:uterm}\ee
in which the exponents must be positive integers or zero. 
Invariance under the three charges yields
\be
n^{(1)}_{ij}=0\ ,\qquad
n^{(2)}_{ij}=Y^{(2)~[u]}_{ij}\ ,\qquad
n^{(3)}_{ij}=-Y^{(1)~[u]}_{ij}+Y^{(2)~[u]}_{ij}\ ,
\ee
where $Y^{(1)~[u]}_{ij}$, and $Y^{(2)~[u]}_{ij}$ are the charges of 
${\bf Q}^{}_i\bar{\bf u}^{}_jH^{}_u$, respectively. They are determined 
by our choice for the charges $Y^{(1,2)}$. A straightforward computation 
yields the orders of magnitude in the charge 
$2/3$ Yukawa matrix 
 \be
Y_{}^{[u]}\sim\left( \begin{array}{ccc}
\lambda^8 &\lambda^5&\lambda^3\\ \lambda^7&\lambda^4&\lambda^2\\
\lambda^5&\lambda^2&1\end{array}  \right)\ ,\ee
where $\lambda=\vert\theta_a\vert/ M$ is the expansion parameter.

A similar computation is now applied to the charge $-1/3$ Yukawa standard 
model invariants ${\bf Q}^{}_i\bar{\bf d}^{}_jH^{}_d$. The difference is 
the absence of dimension-three terms, so that its $X$-charge, which we 
denote by $X^{[d]}$ need not vanish. We find that if 
$X^{[d]}> -3$, one exponent in the $(33)$ position is negative, resulting 
in a supersymmetric zero~\cite{LNS}, and spoiling the quark hierarchy. 
Hence, as long as $X^{[d]}\le -3$, we deduce the charge $-1/3$
Yukawa matrix  

\be 
Y_{}^{[d]}\sim\lambda_{}^{-3X^{[d]}-6}\left( \begin{array}{ccc}
\lambda_{}^{4} &\lambda_{}^{3}&\lambda_{}^{3}\\ 
\lambda_{}^{3}&\lambda_{}^{2}&\lambda_{}^{2}\\
\lambda_{}^{}&1&1\end{array}  \right)\ ,\ee
and diagonalization of the two Yukawa matrices yields the 
CKM matrix
\be
{\cal U}^{}_{CKM}\sim\left( \begin{array}{ccc}
1 &\lambda_{}^{}&\lambda_{}^{3}\\ \lambda_{}^{}&1&\lambda_{}^{2}\\
\lambda_{}^{3}&\lambda_{}^{2}&1\end{array}  \right)\ .\ee
This shows  the expansion parameter to be of the same order of magnitude 
as the Cabibbo angle $\lambda_c$. For definiteness in what follows we 
take them to be equal, although as we show later, the Green-Schwarz evaluation of $\lambda$ 
gives a slightly higher value.

The eigenvalues of these matrices reproduce the 
geometric interfamily hierarchy for quarks of both charges

\be
{m_u\over m_t}\sim \lambda_c^8\ ,\qquad {m_c\over m_t}\sim
\lambda_c^4\ .\ee
\be
{m_d\over m_b}\sim\lambda_c^4\ ,\qquad {m_s\over m_b}\sim \lambda_c^2\
,\ee
while the quark intrafamily hierarchy is given by
\be
{m_b\over m_t}= \cot\beta\lambda_{c}^{-3X^{[d]}-6}\ .\ee
implying the relative suppression of the bottom to  top quark masses, 
without large $\tan\beta$. 
These quark-sector results are the same as in a previously published 
model~\cite{EIR}, but our present model is different in the lepton 
sector. 

The analysis is much the same as for the down quark sector. No 
dimension-three term appears and the standard model invariant 
$L_i\overline e_jH_d$ have charges $X^{[e]}$, $Y_{ij}^{(1,2)~[e]}$.
The pattern of eigenvalues depends on the $X^{[e]}$: 
if $X^{[e]}>-3$, we find a supersymmetric zero in the $(33)$ position, 
and the wrong hierarchy for lepton masses; if $X^{[e]}=-3$, there are 
supersymmetric zeros in the $(21)$ and $(31)$ position, yielding

\be 
Y_{}^{[e]}\sim\lambda_{c}^{3}\left( \begin{array}{ccc}
\lambda_c^{4} &\lambda_c^{5}&\lambda_c^{3}\\ 
0&\lambda_c^{2}&1\\
0&\lambda_c^2&1\end{array}  \right)\ .\ee
Its diagonalization yields the lepton interfamily hierarchy

\be
{m_e\over m_\tau}\sim\lambda_c^4\ ,\qquad {m_\mu\over m_\tau}\sim
\lambda_c^2\ .\ee
Our choice of $X$ insures that $X^{[d]}=X^{[e]}$, which guarantees 
through the anomaly conditions the 
correct value of the Weinberg angle at cut-off, since 

\be \sin^2\theta_w={3\over 8}~~~\leftrightarrow ~~~X^{[d]}=X^{[e]}\ ;\ee
it  sets $X^{[d]}=-3$, so that  

\be
{m_b\over m_\tau}\sim 1\ ;\qquad {m_b\over m_t}\sim 
\cot\beta\lambda_c^3\ .\ee
It is a remarkable feature of this type of model that both inter- and 
intra-family hierarchies are linked not only with one another but with 
the value of the Weinberg angle as well. In addition, the model predicts
a natural suppression of $m_b/m_\tau$, which suggests that $\tan \beta$
is of order one.


\mysection{Neutrino Masses}

Our model, based on $E_6$, has all the features of $SO(10)$; in particular,
neutrino masses are naturally generated by the seesaw mechanism~\cite{SEESAW}
if the three right-handed neutrinos $\overline N_i$ acquire a Majorana mass
in the DSW vacuum. The flat direction analysis then indicates that their
$X$-charges must be negative half-odd integers, that is
$X_{\overline N}=-1/2,~-3/2,\dots$. 

Their standard-model invariant masses are generated by terms of the form

\be 
M\overline N_i\overline N_j
{\bigl ( {\theta_1 \over M} \bigr )}^{p^{(1)}_{ij}}
{\bigl ( {\theta_2 \over M} \bigr )}^{p^{(2)}_{ij}}
{\bigl ( {\theta_3 \over M} \bigr )}^{p^{(3)}_{ij}}\ ,\ee
where $M$ is the cut-off of the theory. In the $(ij)$ matrix element, 
the exponents are computed to be equal to $-2X_{\overline N}$ plus 

\be
\left( \begin{array}{ccc}
(0,4,0)&(0,2,1)&(0,0,-1)\\ (0,2,1)&(0,0,2)&(0,-2,0)\\ 
(0,0,-1)&(0,-2,0)&(0,-4,-2)
\end{array}  \right)\ ,\ee
If $X_{\overline N}=-1/2$, this matrix has supersymmetric 
zeros in the $(23)$, $(32)$ and $(33)$ elements. While this does not
result in a zero eigenvalue, the absence of these invariants from the
superpotential creates flat directions along which $\langle \overline
N_3 \rangle \neq 0$; such flat directions are dangerous because they
 can lead to vacua other than the DSW vacuum. If $X_{\overline N} \leq -5/2$,
none of the entries of the Majorana mass matrix vanishes; but then the vacuum
analysis indicates that flat directions are allowed which involve MSSM fields.
For those reasons, we choose $X_{\overline N}=-3/2$, which 
still yields one harmless  supersymmetric zero in the Majorana 
mass matrix, now of the form

\be
M\lambda_c^{7}\left( \begin{array}{ccc}
\lambda_c^{6} &\lambda_c^{5}&\lambda_c^{}\\ 
\lambda_c^{5}&\lambda_c^{4}&1\\
\lambda_c^{}&1&0\end{array}  \right)\ .\ee
Its diagonalization yields three massive right-handed neutrinos with masses 

\be
m_{\overline N_e}\sim M\lambda_c^{13}\ ;\qquad m_{\overline N_\mu}\sim 
m_{\overline N_\tau}\sim M\lambda_c^7\ .\ee

By definition, right-handed neutrinos are those that 
couple to the standard-model invariant $L_iH_u$, 
and serve as Dirac partners to the chiral neutrinos. In our model,

\be
X(L_iH_u\overline N_j)\equiv X^{[\nu]}=0\ .\ee
The superpotential contains the terms 

\be
L_iH_u\overline N_j
{\bigl ( {\theta_1 \over M} \bigr )}^{q^{(1)}_{ij}}
{\bigl ( {\theta_2 \over M} \bigr )}^{q^{(2)}_{ij}}
{\bigl ( {\theta_3 \over M} \bigr )}^{q^{(3)}_{ij}}\ \ee
resulting, after electroweak symmetry breaking, in the orders of magnitude
(we note $v_u = \vev{H^0_u}$)

\be
 v_u  \left( \begin{array}{ccc}
\lambda_c^{8}&\lambda_c^{7}&\lambda_c^{3}\\  \lambda_c^{5}&\lambda_c^{4}&1
\\ \lambda_c^{5}&\lambda_c^{4}&1\end{array}  \right)\ \ee
for the neutrino Dirac mass matrix. The actual 
neutrino mass matrix is generated by the seesaw mechanism. 
A careful calculation yields the orders of magnitude

\be
{v_u^2\over M\lambda_c^3} \left( \begin{array}{ccc}
\lambda_c^6&\lambda_c^3&\lambda_c^3\\ \lambda_c^3&1&1\\ \lambda_c^3&1&1
\end{array}  \right)\ .
\label{eq:nu_matrix}\ee
A characteristic of the seesaw mechanism is that the 
charges of the $\overline N_i$ do not enter in the determination of these 
orders of magnitude as long as there are no massless right-handed
neutrinos. Hence the structure of the neutrino mass matrix depends only on
the charges of the invariants $L_iH_u$, already fixed by phenomenology and
anomaly cancellation. In the few models with two non-anomalous horizontal
symmetries based on $E_6$ that reproduce the observed quark and charged
lepton masses and mixings, the neutrino mass spectrum exhibits the same
hierarchical structure: the matrix (\ref{eq:nu_matrix}) is a very stable
prediction of our model. Its diagonalization yields the neutrino mixing
matrix~\cite{MNS}

\be
 {\cal U}_{\rm MNS}=\left( \begin{array}{ccc}
1&\lambda_c^{3}&\lambda_c^{3}\\  \lambda_c^{3}&1&1
\\ \lambda_c^{3}&1&1\end{array}  \right)\ ,\ee
so that the mixing of the electron neutrino is small, of the order of
$\lambda_c^3$, while the mixing between the $\mu$ and $\tau$ neutrinos is of
order one. Remarkably enough, this mixing pattern is precisely the one
suggested by the non-adiabatic MSW ~\cite{MSW} explanation of the solar
neutrino deficit and by the oscillation interpretation of the reported
anomaly in atmospheric neutrino fluxes (which has been recently confirmed
by the Super-Kamiokande \cite{SuperK} and Soudan \cite{Soudan}
collaborations). It should be stressed here that the model of Ref.
\cite{EIR}, which differs from the present one by the fact that $Y^{(1)}$
is along $B+L$ instead of $B-L$, predicts the same lepton mixing matrix.
However, it cannot accomodate the MSW effect, because it yields an inverted
mass hierarchy in the neutrino sector. The change of $B+L$ into $B-L$
restores the natural hierarchy, but requires the addition of vector-like
matter to cancel anomalies.

Whether the present model actually fits the experimental data on solar and
atmospheric neutrinos or not depends on the eigenvalues of the mass matrix
(\ref{eq:nu_matrix}). A naive order of magnitude diagonalization gives a
$\mu$ and $\tau$ neutrinos of comparable masses, and a much lighter electron
neutrino:
\be
m_{\nu_e}\ \sim\ m_0\, \lambda_c^{6}\ ;\qquad 
m_{\nu_\mu},\, m_{\nu_\tau}\ \sim\ m_0\ ;\qquad
m_0\ =\ {v_u^2\over M\lambda_c^3}\ ,
\label{eq:nu_mass}\ee
The overall neutrino mass scale $m_0$ depends on the cut-off $M$. Thus the
neutrino sector allows us, in principle, to measure it.

At first sight, this spectrum is not compatible with a simultaneous
explanation of the solar and atmospheric neutrino problems, which requires
a hierarchy between $m_{\nu_\mu}$ and $m_{\nu_\tau}$. However, the estimates
(\ref{eq:nu_mass}) are too crude: since the (2,2), (2,3) and (3,3) entries
of the mass matrix all have the same order of magnitude, the prefactors that
multiply the powers of $\lambda_c$ in (\ref{eq:nu_matrix}) can spoil the
naive determination of the mass eigenvalues. In order to take this effect
into account, we rewrite the neutrino mass matrix, expressed in the basis
of charged lepton mass eigenstates, as:

\be
m_0\, \left( \begin{array}{ccc}
a \lambda_c^6 & b \lambda_c^3 & c \lambda_c^3 \\ b \lambda_c^3 & d & e \\
c \lambda_c^3 & e & f
\end{array}  \right)\ ,\ee
where the prefactors $a$, $b$, $c$, $d$, $e$ and $f$, unconstrained by any
symmetry, are assumed to be of order one, say $0.5 < a, \ldots f < 2$.
Depending on their values, the two heaviest neutrinos may be either
approximately degenerate (scenario 1) or well separated in mass (scenario 2).
It will prove convenient in the following discussion to express their mass
ratio and mixing angle in terms of the two parameters
$x = \frac{df-e^2}{(d+f)^2}$ and $y = \frac{d-f}{d+f}$:

\be
\frac{m_{\nu_2}}{m_{\nu_3}}\ =\ \frac{1-\sqrt{1-4x}}{1+\sqrt{1-4x}}\ ;
\qquad \sin^2 2 \theta_{\mu \tau}\ =\ 1\ -\ \frac{y^2}{1-4x}\ .
\label{eq:mixing_mu_tau} \ee
Scenario 1 corresponds to both regimes $4x \sim 1$ and $(-4x) \gg 1$, while
scenario 2 requires $|x| \ll 1$. Let us stress that small values of $|x|$
are very generic when $d$ and $f$ have same sign, provided that
$df \sim e^2$. Since this condition is very often satisfied by arbitrary
numbers of order one, a mass hierarchy is not less natural, given the
structure (\ref{eq:nu_matrix}), than an approximate degeneracy.

{\bf Scenario 1: $m_{\nu_2} \sim m_{\nu_3}$}. In this scenario, the
oscillation frequencies $\Delta m^2_{ij} = m^2_{\nu_j} - m^2_{\nu_i}$ are
roughly of the same order of magnitude, $\Delta m^2_{12} \sim \Delta m^2_{23}
\sim \Delta m^2_{13}$. There is no simultaneous explanation of the solar and
atmospheric neutrino data. A strong degeneracy between $\nu_2$ and $\nu_3$,
which would result in two distinct oscillation frequencies, $\Delta m^2_{23}
\ll \Delta m^2_{12} \simeq \Delta m^2_{13}$, would be difficult to achieve
in this model\footnote{This is to be contrasted with the models of Ref.
\cite{BLPR}, in which the close degeneracy is linked to the structure of the
neutrino mass matrix.}, as it would require either to fine-tune $d \simeq f$
and to allow for $e \ll 1$ (case $4x \sim 1$), or to fine-tune $d \simeq - f$
(case $(-4x) \gg 1$).

Thus, this scenario yields only the MSW effect, with $\Delta m^2_{12}
\sim \Delta m^2_{13} \sim 10^{-6}\, eV^2$, and a total electron neutrino
oscillation probability
\be
P\, (\nu_e \rightarrow \nu_{\mu, \tau})\ =\ 4\, u^2
\lambda_c^6 \sin^2 \left( \frac{\Delta m^2_{12} L}{4 E} \right)\
+\ 4\, v^2 \lambda_c^6
\sin^2 \left( \frac{\Delta m^2_{13} L}{4 E} \right)\
\label{eq:nu_e} ,\ee
where the parameters $u$ and $v$ are defined to be $u =\frac{bf-ce}{df-e^2}$
and $v =\frac{be-cd}{df-e^2}\ $. If $\Delta m^2_{12}$ is close enough to
$\Delta m^2_{13}$, (\ref{eq:nu_e}) can be viewed as a two-flavour
oscillation with a mixing angle $\sin^2 2 \theta = 4\, (u^2+v^2)\,
\lambda_c^6$. The solar neutrino data then require $(u^2+v^2) \sim 10-20$
\cite{Langacker}, which is still reasonable in our approach. Although the
mixing between $\mu$ and $\tau$ neutrinos is of order one, they are too
light to account for the atmospheric neutrino anomaly.

{\bf Scenario 2: $m_{\nu_2} \ll m_{\nu_3}$}. The two distinct oscillation
frequencies $\Delta m^2_{12}$ and $\Delta m^2_{13} \simeq \Delta m^2_{23}$
can explain both the solar and atmospheric neutrino data: non-adiabatic MSW
$\nu_e \rightarrow \nu_{\mu, \tau}$ transitions require \cite{Langacker}

\be
 4 \times 10^{-6}\, eV^2\ \leq\ \Delta m^2\ \leq\ 10^{-5}\, eV^2
\qquad (\mbox{best fit:}\ 5 \times 10^{-6}\: eV^2)\ ,\ee
while an oscillation solution to the atmospheric neutrino anomaly requires
\cite{atmospheric}

\be
 5 \times 10^{-4}\, eV^2\ \leq\ \Delta m^2\ \leq\ 5 \times 10^{-3}\, eV^2
\qquad (\mbox{best fit:}\ 10^{-3}\: eV^2)\ .\ee
To accommodate both, we need $0.03 \leq \frac{m_{\nu_2}}{m_{\nu_3}} \simeq x
\leq 0.15$ (with $x=0.06$ for the best fits), which can be achieved without
any fine-tuning in our model. Interestingly enough, such small values of $x$
generically push $\sin^2 2 \theta_{\mu \tau}$ towards its maximum, as can be
seen from (\ref{eq:mixing_mu_tau}). Indeed, since $d$ and $f$ have the same
sign and are both of order one, $y^2$ is naturally small compared with
$(1-4x)$. This is certainly a welcome feature, since the best fit to the
atmospheric neutrino data is obtained precisely for $\sin^2 2 \theta = 1$.

To be more quantitative, let us fix $x$ and try to adjust $y$ to make
$\sin^2 2 \theta_{\mu \tau}$ as close to $1$ as possible. With $x=0.06$,
one obtains $\sin^2 2 \theta_{\mu \tau} = 0.9$ for $y \simeq 0.3$,
$\sin^2 2 \theta_{\mu \tau} = 0.95$ for $y \simeq 0.2$ and
$\sin^2 2 \theta_{\mu \tau} = 0.98$ for $y \simeq 0.1$. This shows that
very large values of $\sin^2 2 \theta_{\mu \tau}$ can be obtained without
any fine-tuning (note that $y=1/3$ already for $d/f=2$). Thus, in the
regime $x \ll 1$, $\nu_\mu \leftrightarrow \nu_\tau$ oscillations provide
a natural explanation for the observed atmospheric neutrino anomaly. As
for the solar neutrino deficit, it can be accounted for by MSW transitions
from the electron neutrinos to both $\mu$ and $\tau$ neutrinos, with
parameters $\Delta m^2 = \Delta m^2_{12}$ and $\sin^2 2 \theta = 4\, u^2
\lambda^6$. To match the mixing angle with experimental data, one needs
$u \sim 3-5\, $; we note that such moderate values of $u$ are favoured by
the fact that $df \sim e^2$. 

In both scenarios, the scale of the neutrino masses measures the cut-off
$M$. In scenario 1, the MSW effect requires $m_0 \sim 10^{-3}\, eV$, which
gives $M \sim  10^{18}\, GeV$. In scenario 2, the best
fit to the atmospheric neutrino data gives $ m_0\, (d+f) = m_{\nu_2}
+ m_{\nu_3} \simeq 0.03\, eV$, which corresponds to a slightly lower 
cut-off, $10^{16}\, GeV \leq M \leq 4 \times 10^{17}\, GeV$
(assuming $0.2 \leq d+f \leq 5$). It is remarkable that those values are
so close to the unification scale obtained by running the standard 
model gauge couplings. This result depends of course on our choice
for $X_{\overline N}$, since
\be
m_0 = {v_u^2\over M}\ \lambda_c^{6(1+X_{\overline N})} ,\ee
but the value $X_{\overline N} = -3/2$ is
precisely that favored by the flat direction analysis. As a comparison,
$X_{\overline N}=-1/2$ would give $M \sim 10^{22}\, GeV$, and
$X_{\overline N} \leq -5/2$ corresponds to $M < 10^{14}\, GeV$.

Turning the argument the other way, had we set $M = M_{U}$ {\it ab initio},
the value of $X_{\overline N}$ favored by the flat direction analysis would
yield precisely the neutrino mass scale needed to explain the solar neutrino
deficit, $m_0 \sim 10^{-3}\, eV$. Other values of $X_{\overline N}$ would
give mass scales irrelevant to the data: $X_{\overline N}=-1/2$ corresponds
to $m_0 \sim 10^{-7} ~{\rm eV}$, which is not interesting for neutrino
phenomenology, and $X_{\overline N} \leq -5/2$ to $m_0 > 10 ~{\rm eV}$,
which, given the large mixing between $\mu$ and $\tau$ neutrinos (and
assuming no fine-tuned degeneracy between them), is excluded by oscillation
experiments.

To conclude, our model can explain both the solar neutrino deficit and the
atmospheric neutrino anomaly, depending on the values of the order-one
factors that appear in the neutrino mass matrices. The cut-off $M$, which
is related to the neutrino mass scale, is determined to be close to the
unification scale. Finally, our model predicts neither a neutrino mass in the
few $eV$ range, which could account for the hot component of the dark matter
needed to understand structure formation, nor the LSND result \cite{LSND}.
The upcoming flood of experimental data on neutrinos will severely test
our model.


\mysection{Vector-like Matter}
\label{section:VL}

To cancel 
anomalies involving hypercharge, vector-like matter with standard-model 
charges must be present. Its nature  is not fixed by 
phenomenology, but by a variety of theoretical requirements: vector-like 
matter  
must not affect the unification of gauge couplings, must cancel 
anomalies, must yield the value of the Cabibbo angle,  must not create 
unwanted flat directions in the MSW vacuum, and of course must be sufficiently 
massive to have avoided detection. As we shall see below, our 
$E_6$-inspired  model, with vector-like matter in $\bf 5-\overline 5$ 
combinations, comes close to satisfying these 
requirements, except that it produces a high value for the expansion 
parameter.

The masses of the three families of standard 
model vector-like matter
are determined through the same procedure, namely operators of the
form

\be
M\overline{\bf D}_i{\bf D}_j
{\bigl ( {\theta_1 \over M} \bigr )}^{s^{(1)}_{ij}}
{\bigl ( {\theta_2 \over M} \bigr )}^{s^{(2)}_{ij}}
{\bigl ( {\theta_3 \over M} \bigr )}^{s^{(3)}_{ij}}
+M\overline E_iE_j
{\bigl ( {\theta_1 \over M} \bigr )}^{t^{(1)}_{ij}}
{\bigl ( {\theta_2 \over M} \bigr )}^{t^{(2)}_{ij}}
{\bigl ( {\theta_3 \over M} \bigr )}^{t^{(3)}_{ij}}\ .
\ee  
The $X$-charges of the standard model invariant 
mass terms are the same 
\be
X(\overline{\bf D}_i{\bf D}_j)=X(\overline E_iE_j)=
2\alpha-4\gamma\equiv -{\bf n}.\ee
Its value determines the $X$-charge, since 
$X^{[d]}=-3$ and $X_{\overline
N_i}=-3/2$ already fix  $\beta=-3/20$ and $\alpha+\gamma=-3/4$. 
It also fixes the orders of magnitude of the vector-like masses.
\par First we note that ${\bf n}$ must be a non-negative integer. The
reason is that the power of $\theta _1$ is ${\bf n}$, the X-charge of
the invariant and by holomorphy, it must be zero or a positive
integer. Thus if ${\bf n}$ is negative, all vector-like matter is massless, 
which is not acceptable. 
The exponents for the heavy quark matrix are given by the integer
${\bf n}$ plus
\be
\left( \begin{array}{ccc}
(0,-3,-3)&(0,-1,-3)&(0,1,-1)\\ (0,-2,0)&(0,0,0)&(0,2,2)\\ 
(0,-1,1)&(0,1,1)&(0,3,3)
\end{array}  \right)\;\;\;\;~:\;\;\overline{\bf D}_i{\bf D}_j\ .\ee
Those of the heavy leptons, by ${\bf n}$ plus
\be
\left( \begin{array}{ccc}
(0,-3,-3)&(0,-2,-2)&(0,-1,-1)\\ (0,-1,-1)&(0,0,0)&(0,1,1)\\ 
(0,1,1)&(0,2,2)&(0,3,3)
\end{array}  \right)\;\;\;\;~:\;\;\overline E_iE_j\ .\ee
  
\par Since these particles carry 
standard model quantum numbers, they can 
affect gauge coupling unification. 
As these states fall into complete $SU(5)$ representations, the gauge
couplings unify at one loop like in the MSSM, provided that the mass splitting 
between
the doublet and the triplet is not too large. 
\par $\bullet~$ ${\bf n}=0$.  We obtain the mass matrices
\be M_{{\bf \overline D}{\bf D}}=M\pmatrix {0&0&0 \cr 0&1&\lambda _c^4 \cr 
0&\lambda _c^2&\lambda _c^6\cr}\;\;\;\;~,\;\;\;\;~
    M_{{\overline E}{E}}=M\pmatrix {0&0&0 \cr 0&1&\lambda _c^2 \cr 
\lambda _c^2&\lambda _c^4&\lambda _c^6\cr}
\ee
Diagonalization of these matrices yields one zero eigenvalue for both
matrices and nonzero (order of magnitude) eigenvalues 
$M$ and 
$\lambda _c^7M$ for $M_{{\bf \overline D}{\bf D}}$ and 
$M$ and $\lambda _c^2M$ for $M_{{\overline E}{E}}$. The 
pair of zero eigenvalues is clearly undesirable and furthermore the mass 
splitting
between the second family $E$ and ${\bf D}$ destroys gauge
coupling unification.  
This excludes ${\bf n}=0$.
\par  $\bullet~$ ${\bf n}=1$. The mass matrices are
\be M_{{\bf \overline D}{\bf D}}=M\pmatrix {0&0&\lambda _c^3 \cr 0&\lambda 
_c^3&\lambda _c^7 \cr 
\lambda _c^3&\lambda _c^5&\lambda _c^9\cr}\;\;\;\;~,\;\;\;\;~
    M_{{\overline E}{E}}=M\pmatrix {0&0&\lambda _c \cr \lambda _c&\lambda 
_c^3&\lambda _c^5 \cr 
\lambda _c^5&\lambda _c^7&\lambda _c^9\cr}.
\ee
The eigenvalues for $M_{{\bf \overline D}{\bf D}}$ are $\lambda _c^3M$,
$\lambda _c^3M$ and $\lambda _c^3M$ and for $M_{{\overline E}{E}}$   
$\lambda _c M$, $\lambda _c M$ and $\lambda _c^9M$. The splitting between the
members of the third
family vector-like fields is too large and as a consequence, gauge coupling 
unification is spoiled.  
\par  $\bullet~$ ${\bf n}=2$. The mass matrices are:
\be M_{{\bf \overline D}{\bf D}}=M\pmatrix {0&0&\lambda _c^6 \cr 
\lambda _c^4&\lambda _c^6&\lambda _c^{10} \cr 
\lambda _c^6&\lambda _c^{8}&\lambda _c^{12}\cr}\;\;\;\;~,\;\;\;\;~
    M_{{\overline E}{E}}=M\pmatrix {0&\lambda _c^2&\lambda _c^4 \cr \lambda 
_c^4&\lambda _c^6&\lambda _c^{8} \cr 
\lambda _c^8&\lambda _c^{10}&\lambda _c^{12}\cr}.
\ee
The eigenvalues are now 
$\lambda _c^4M$, $\lambda _c^6M$, $\lambda _c^{8}M$ and $\lambda _c^2M$, 
$\lambda _c^4M$,
$\lambda _c^{12}M$, respectively. There is again splitting between the
families of the doublet and the triplet and therefore the gauge
couplings do not unify at one loop. The splitting in this case is not too
big and a two loop analysis may actually prove this case viable from
the gauge coupling unification point of view.

\par $\bullet~$ ${\bf n}=3$. We obtain the 
mass matrices
\be M_{{\bf \overline D}{\bf D}}=M\pmatrix {\lambda _c^3&\lambda _c^5&\lambda 
_c^9 \cr 
\lambda _c^7&\lambda _c^9&\lambda _c^{13} \cr \lambda _c^9
&\lambda _c^{11}&\lambda _c^{15}\cr}\;\;\;\;~,\;\;\;\;~
    M_{{\overline E}{E}}=M\pmatrix {\lambda _c^3&\lambda _c^5&\lambda _c^{7} 
\cr 
\lambda _c^7&\lambda _c^9&\lambda _c^{11} \cr 
\lambda _c^{11}&\lambda _c^{13}&\lambda _c^{15}\cr}
\ee  
with eigenvalues:
\be M_{\bf D}=\{\lambda _c^3M,\;\; \lambda _c^9M,\;\; \lambda
_c^{15}M\}\;\;\ee 
and 
\be M_{E}=\{\lambda _c^3M,\;\; \lambda _c^9M,\;\; \lambda _c^{15}M\}\ ,\ee  
respectively. The unification of couplings 
in this case is preserved. For ${\bf n}\ge 3$, 
there are no supersymmetric zeros in the mass matrices
and the mass eigenvalues are just the diagonal entries, so  
there is no splitting between masses of the same family of ${\bf D}$ and $E$. 
A simple one-loop analysis using self-consistently $M=M_U$ in the mass 
of the vector-like particles and for the unification scale, yields unified  
gauge couplings at the 
unification scale, $M_U$
\be
\alpha(M_U)\sim{1\over 19}\ ;\qquad M_U\sim 3\times 10^{16} {\rm GeV}\ .\ee

For ${\bf n}$ large, other problems arise as the vector-like matter 
becomes too light. This can easily spoil gauge coupling unification by
two loop effects \cite{Quarksinglets} 
and cause significant deviations from precision measurements of standard model 
parameters~\cite{Precision},\cite{Quarksinglets}. 
Thus the unification of the gauge couplings favors ${\bf n}=3$.

\par The value of ${\bf n}$ also determines the mixing between the chiral
and vector-like matter. Indeed, the quantum numbers of the vector-like
matter allow for mixing with the 
chiral families, since ($E_i$, $L_j$, $H_d$), (${\overline E}_i$ with
$H_u$) and  
($\overline{\bf D}_i$ with $\overline{\bf d}_j$) have the same standard
model quantum numbers. This generates new standard model
invariants. In table 1, we give a set of  
mixed operators up to superfield dimension 3. Next to the operator we show its 
$X$-charge in brackets.  One notices that the operators fall into three 
classes.

\begin{table}
\vskip -2cm
\caption{{\bf Operators that mix MSSM fields with vector-like matter
with $\beta =-3/20$. }}
\vskip 0.5cm
\hskip 2cm
\begin{center}
\begin{tabular}{|c|c|c|}

           \hline  & &   \\
$~{\rm {\bf Class\; 1}}~$ & $ {\bf Class\; 2} $ & $~{\bf Class\; 3}~$\\ 
                       \hline \hline   & &   \\

$EH_u\;\;[{3\over 2}-{{\bf n}\over 2}]$& 
$L{\overline E}\;\;[-{{\bf n}\over 2}]$&
${\bf \overline u}{\bf \overline D}{\bf \overline D}\;\;[-{\bf n}-{3\over 2}]$ 
\\
                                      & &   \\ 
                                                                 
${\overline E}H_d\;\;[-{3\over 2}-{{\bf n}\over 2}]$& 
${\bf D}{\bf \overline d}\;\;[-{{\bf n}\over 2}]$ & 
$E{\bf Q}{\bf \overline D}\;\;[-{\bf n}-{3\over 2}]$\\ 
                                      & &   \\
${\bf Q}{\bf Q}{\bf D}\;\;[-{3\over 2}-{{\bf n}\over 2}]$& 
${\bf Q}{\bf \overline D}H_d\;\;[-3-{{\bf n}\over 2}]$ & 
$EE{\overline e}\;\;[-{\bf n}-{3\over 2}]$\\ 
                                      & &   \\

${\bf \overline u}{\bf \overline d}{\bf \overline D}\;\;[-{3\over 2}-{{\bf 
n}\over 
2}]$& 
$E\overline eH_d\;\;[-3-{{\bf n}\over 2}]$ & 
$$\\ 
                                      & &   \\

${\bf Q}{\bf \overline u}{\overline E}\;\;[-{3\over 2}-{{\bf n}\over 2}]$& 
$$ & 
$$\\ 
                                      & &   \\

$E{\bf Q}{\bf \overline d}\;\;[-{3\over 2}-{{\bf n}\over 2}]$& 
$$ & 
$$\\ 
                                      & &   \\

$L {\bf Q}{\bf \overline D}\;\;[-{3\over 2}-{{\bf n}\over 2}]$& 
$$ & 
$$\\ 
                                      & &   \\
$LE{\overline e}\;\;[-{3\over 2}-{{\bf n}\over 2}]$& 
$$ & 
$$\\ 
                                      & &   \\

${\bf D}{\bf \overline u}{\overline e}\;\;[-{3\over 2}-{{\bf n}\over 2}]$& 
$$ & 
$$\\ 
 & &   \\

\hline

\end{tabular}
\end{center}
\end{table}

For ${\bf n}$ odd only the operators of the 
first class can appear in the superpotential and for ${\bf n}$ even only 
operators of the second 
class appear. The third class is excluded for any integer
value of ${\bf n}$. Let us examine these two possibilities in more detail.
\par ${\bf n}=2,4,6,...$ Only operators of the
second class are allowed in $W$ and ${\bf \overline D}$ mixes with
${\bf \overline d}$. The mixing is computed by diagonalizing the down type
quark mass matrices. To see this,
we give a one family example where the operators ${\bf \overline
D}{\bf D}$, ${\bf Q}{\bf \overline
d}H_d$, ${\bf Q}{\bf \overline D}H_d$, and ${\bf D}{\bf \overline d}$ are all
present in the superpotential. After electroweak breaking the masses
of the down type quark fields come from diagonalizing the matrix
\be \pmatrix {v_dY^{[d]}&v_dY^{[D]}\cr M_{{\bf D}{\bf \overline
d}}&M_{{\bf \overline D}{\bf D}}\cr }.\ee 
The extra quark fields affect the down quark mass matrices 
of section 5 and modify our previous order of magnitude estimates. 
The same type of mixing happens in the lepton sector due to the operators 
${\overline E}E$, $L{\overline E}$ and $E{\overline e}H_d$. 
If allowed, this type of mixing produces phenomenologically
unacceptable mass patterns for quarks and charged leptons.   
\par ${\bf n}=3,5,..$. Operators of the first
class are allowed since their $X$ charges are
all negative integers. Due to the mixing of the heavy leptons with
the Higgs doublets, we have to diagonalize the following mass matrix
(we give again a simple one family example):
\be \pmatrix {\mu &M_{{\overline E}H_d}\cr 
              M_{EH_u}&M_{{\overline E}E}\cr}.\ee
The 11 entry is the $\mu $ term generated by the Giudice-Masiero
mechanism and is naturally of order of a TeV. 
The Higgs eigenstates will be modified to 
\be H_u'=H_u+\sum _i{c^u_i\cdot {\overline E}_i}\ee
and 
\be H_d'=H_d+\sum _i{c^d_i\cdot {E}_i}\ee
where $c^{u,d}_i$ are mixing angles to be obtained upon
diagonalization. 
With both off diagonal entries present, this matrix has two large eigenvalues
and consequently the Higgs mass is
driven to the Planck scale. If one of the off diagonal entries is
missing, then the matrix has one small and one large eigenvalue and the
mixing is harmless as long as the angles $c^{u,d}_i$ are small (see later). 

\par There are several ways to evade these problems. One is 
to relax the simple but
very restrictive assumption that $X$ is the same for both
the MSSM and the vector-like fields and another is
to assume
the existence of a discrete symmetry that prohibits the dangerous
operators.

\subsection{Shift X.}
The vector-like matter could come from a different ${\bf 27}$ than the
MSSM fields so that the $X$-charges of the vector-like fields are shifted 
relative to the fields in the
${\bf 16}$: 
\be X_{VL}={\overline \alpha} +{\overline \beta}V +{\overline \gamma}V'.\ee

In table 2 we show the different operators with their
$X$-charges. It is interesting to notice that the $X$ charges
of these operators depend only on ${\overline \beta}$ and ${\bf
n}=-2{\overline \alpha}+4{\overline \gamma}$. We have two possibilities.

\begin{table}
\vskip -2cm
\caption{{\bf Operators that mix MSSM fields with vector-like matter
with $X_{VL}={\overline \alpha}+{\overline \beta}V+{\overline \gamma}V'$. }}
\vskip 0.5cm
\hskip -1.1cm
\begin{center}
\begin{tabular}{|c|c|c|}

           \hline  & &   \\
$~{\rm {\bf Class\; 1}}~$ & $ {\bf Class\; 2} $ & $~{\bf Class\; 3}~$\\ 
                       \hline \hline   & &   \\

$EH_u\;\;[-(-2{\overline \beta}+{{\bf n}\over 2}-{18\over 10})]$& 
$L{\overline E}\;\;[(-2{\overline \beta}+{{\bf n}\over 2}+{3\over 10})]$&
${\bf \overline u}{\bf \overline D}{\bf \overline D}\;\;[-(-4{\overline 
\beta}+{{\bf n}}+{9\over 10})]$ 
\\ & &   \\ 
${\overline E}H_d\;\;[-(2{\overline \beta}+{{\bf n}\over 2}+{18\over 10})]$& 
${\bf D}{\bf \overline d}\;\;[-(2{\overline \beta}+{{\bf n}\over 2}+{3\over 
10})]$ & 
$E{\bf Q}{\bf \overline D}\;\;[-(-4{\overline \beta}+{{\bf n}}+{9\over 10})]$\\ 
                                      & &   \\
${\bf Q}{\bf Q}{\bf D}\;\;[-(2{\overline \beta}+{{\bf n}\over 2}+{18\over 
10})]$& 
${\bf Q}{\bf \overline D}H_d\;\;[-(-2{\overline \beta}+{{\bf n}\over 2}+{27\over 
10})]$ & 
$EE{\overline e}\;\;[-(-4{\overline \beta}+{{\bf n}}+{9\over 10})]$\\ 
                                      & &   \\

${\bf \overline u}{\bf \overline d}{\bf \overline D}\;\;[-(-2{\overline 
\beta}+{{\bf 
n}\over 
2}+{12\over 
10})]$& 
$E\overline eH_d\;\;[-(-2{\overline \beta}+{{\bf n}\over 2}+{27\over 10})]$ & 
$$\\ 
                                      & &   \\

${\bf Q}{\bf \overline u}{\overline E}\;\;[-(2{\overline \beta}+{{\bf n}\over 
2}+{18\over 
10})]$& 
$$ & 
$$\\ 
                                      & &   \\

$E{\bf Q}{\bf \overline d}\;\;[-(-2{\overline \beta}+{{\bf n}\over 2}+{12\over 
10})]$& 
$$ & 
$$\\ 
                                      & &   \\

$L {\bf Q}{\bf \overline D}\;\;[-(-2{\overline \beta}+{{\bf n}\over 2}+{12\over 
10})]$& 
$$ & 
$$\\ 
                                      & &   \\
$LE{\overline e}\;\;[-(-2{\overline \beta}+{{\bf n}\over 2}+{12\over 10})]$& 
$$ & 
$$\\ 
                                      & &   \\

${\bf D}{\bf \overline u}{\overline e}\;\;[-(2{\overline \beta}+{{\bf n}\over 
2}+{18\over 
10})]$& 
$$ & 
$$\\ 
                                      & &   \\

\hline

\end{tabular}
\end{center}
\end{table}

\subsubsection{No MSSM-Vector Like mixing} We can choose ${\overline \beta}$ in 
such a way that none of
the $X$ charges of the operators appearing in table 2 is an
integer for any integer ${\bf n}$. None of them will appear in $W$ and 
therefore we avoid
the mixing problem. Then, the lightest of the vector-like
fields will be stable. To avoid cosmological problems, this requires a
reheating temperature lower than the lowest vector-like mass in order
to dilute their abundance during inflation. Recall that the mass of 
the 
lightest pair of
${\bf D}$ and $E$ for ${\bf n}=3$ is $\lambda _c ^{15}M\sim 10^{6-7}
 GeV$, and therefore a reheating temperature of at most this order of
magnitude is required. ${\bf n}=4$ or higher result in lower
eigenvalues and thus lower reheating temperatures. 
We therefore favor in this case ${\bf n}=3$. Similar
arguments apply to any other scenario with stable heavy vector-like
states.   
\subsubsection{Partial MSSM-Vector Like mixing} 
Let us take ${\bf n}=3$ which avoids the dangerous  ${\bf 
\overline
d}-{\bf \overline D}$ and ${L}-{\overline E}$ mixing. The $X$ charges of
the operators that could give rise to mixing are $X(EH_u)=2{\overline \beta}
+3/10$ and $X({\overline E}H_d)=-2{\overline \beta}-33/10$. We can choose
${\overline \beta}$ in a way that $X(EH_u)$ is positive
and $X({\overline E}H_d)$ is negative and so prohibit $EH_u$ from
appearing but allow ${\overline E}H_d$. This yields the mass matrix
(7.10) with its 21 element being zero. As we mentioned before, the mixing
is harmless if the angles $c^{u,d}_i$ are small which is indeed the case.  
\par We still have to check if the proton decays due to mixed
operators slowly enough to avoid conflict with experimental
data. Proton decay due to operators consisting only of MSSM fields will
be discussed in a separate section, since it is independent of the
choice of the charges of the vector-like matter. 
\par We find that the dominant proton decay channels come from the operators 
\be L{\bf Q}{\bf \overline D}\;\; {\rm and}\;\; {\bf \overline u}{\bf
\overline d}{\bf \overline D}\ee
and 
\be {\bf Q}{\bf Q}{\bf D}\;\; {\rm and}\;\; {\bf D}{\bf \overline u}{\overline 
e}\ee
via an intermediate heavy quark.
They appear after DSW breaking as
\be \lambda _{ijk}L_i{\bf Q}_j{\bf \overline D}_k+{\overline \lambda}_{ijk}{\bf
\overline u}_{i}{\bf \overline d}_{j}{\bf \overline D}_{k}\ee
and 
\be { \rho} _{ijk}{\bf Q}_i{\bf Q}_j{\bf D}_k+{\overline \rho}_{ijk}{\bf 
D}_{i}{\bf
\overline u}_{j}{\overline e}_{k}\ee
where 
\be \lambda _{ijk}\sim{({<\theta _1>\over M})^{n^{(1)}}}{({<\theta _2>\over 
M})^{n^{(2)}}}{({<\theta _3>\over M})^{n^{(3)}}}\ee
is the suppression factor in the DSW vacuum in front of the
corresponding operator with flavor indices $i$,$j$,$k$. Similar
expressions hold for ${\overline \lambda}_{ijk}$, $\rho _{ijk}$ and
${\overline \rho}_{ijk}$.     
The experimental constraint on these
is \cite{BNir}
\be \lambda _{ijk}{\overline \lambda}_{ijk}\le M_{\bf 
D}^2~10^{-32}~GeV^{-2}\ee 
and similarly
\be \rho _{ijk}{\overline \rho}_{ijk}\le M_{\bf 
D}^2~10^{-32}~GeV^{-2}.\ee
We computed the suppression factors of these operators in the DSW vacuum that 
the model gives for 
${\overline \beta} =7/20$ and we found that the above constraints are
not easily satisfied. Notice that this choice amounts to shifting the $X$-charge of 
the vector-like matter by half a unit of $V$. Interestingly enough, a
similar mechanism 
occurs in some superstring models, as a result of Wilson line breaking 
\cite{Relics}.           
 
\subsection{Discrete symmetry}
It is known that superstring models usually
contain discrete symmetries. If present, they could forbid the 
dangerous mixed operators, leaving the mass terms for the vector-like
matter intact.   
\par As an example, consider the discrete symmetry where  
\be E\rightarrow -E,\;\; {\overline E}\rightarrow -{\overline E}, \;\;{\bf 
D}\rightarrow 
-{\bf D},\;\; {\bf \overline D}\rightarrow -{\bf \overline D}.\ee 
This additional symmetry, indeed completely decouples the MSSM
fields from the vector-like matter. 
No operator with an odd number of vector-like fields is allowed for any value of 
${\bf n}$. Specifically, all operators that mix
MSSM fields and vector-like matter and that can cause
proton decay are also prohibited. Such, are the dimension-3 operators
\be L{\bf Q}{\bf \overline D}\;\;\; {\rm and} \;\;\; {\bf \overline u}{\bf 
\overline 
d}{\bf 
\overline
D}\ee
that belong to class 1 and the dimension-4 operators
\be   {\bf Q}{\bf Q}{\bf Q}E,\;\;\;{\bf \overline u}{\bf \overline u}
{\bf \overline D}{\overline e}.\ee
As a consequence of this discrete symmetry, the vector-like
matter has no available decay channels. This can have undesired 
cosmological implications except if inflation takes place at 
a temperature lower than the lightest of the vector-like particles. 
For this reason we strongly favor the value ${\bf n}=3$.
Also in this case we can keep the simple universal $X$ charge
assignement $X=\alpha +\beta V +\gamma V'$ for both the MSSM and the
vector-like fields which makes the flat direction
analysis particularly simple because the
superpotential has a very small number of
supersymmetric zeros corresponding to standard model invariants 
with vector-like fields.

\par 
\subsection{Summary} To summarize, we have given three alternative ways to fix 
the $X$
charges of the vector-like fields. 
\par ${\bf -}$ The solution of section 7.1.1 is viable
for a reheating temperature $\sim 10^{6-7}\; GeV$ for ${\bf n}=3$.
Lower reheating temperatures are required as ${\bf n}$ increases so in this case
${\bf n}=3$ is clearly favored.   
\par ${\bf -}$ The solution of section 7.1.2 (${\overline \beta
}=7/20$) is not viable even if the mixing angles $c^{u,d}_i$ are small
because the proton decays too fast. The
vector-like particles can decay. 
\par ${\bf -}$ The solution of section 7.2 involves a discrete
symmetry. Stable heavy quarks and leptons require a reheating
temperature $\sim 10^{6-7}\; GeV$ for ${\bf n}=3$ and lower temperatures
for higher values of ${\bf n}$, so ${\bf n}=3$ is again
favored. In this case the flat direction analysis is
particularly simple. 
\par We do not have any physical motivation that can tell us which of
the above proposed mechanisms is the correct one. 
The simplest is the scenario with the discrete symmetry
and from now on we will 
continue our discussion on flat directions and proton decay in this context.


\mysection{The Hidden sector}

So far we have described the  matter  necessary  to satisfy the  anomaly 
conditions that involve standard model quantum numbers, the breaking of 
the extra gauge symmetries,  and phenomenology. These are the three 
chiral families, the three right-handed neutrinos, the three vector-like 
families just described,  and three $\theta $ fields necessary to produce 
the DSW vacuum. We refer to this as {\it visible} matter. By fixing the value of 
$X({\overline E}{E})=X({\bf \overline D}{\bf D})=-{\bf n}$, the $X$ 
charge is totally determined. Since gauge unification favors ${\bf 
n}=3$,  the weak and color anomalies are  fixed,  $C_{\rm color}=C_{\rm 
weak}=-18$.

This enables us to ``predict'' the value of the Cabibbo angle through the relation

\be \lambda _c\sim\lambda=
{<\theta> \over M}=\sqrt{{-g^2_{\rm string}\over {192{\pi ^2}}}
C_{\rm grav}}\ee
Using the Green-Schwarz relation 

\be {{C_{\rm grav}}\over {12}}={{C_{\rm weak}}\over k_{\rm weak}},\ee
and the identification

\be g^2_{\rm string}=k_{\rm weak}g^2_{\rm weak},\ee
we relate the Cabibbo angle to the gauge couplings at the cut-off $\alpha 
(M)$, using only {\it visible} matter contributions

\be \lambda _c\sim\sqrt{{-{C_{\rm weak}\over {4\pi}}}\alpha (M)}\ .\ee
For ${\bf n}=3$, the couplings unify with $\alpha \sim 1/19$, which yields
 $\lambda=0.28$, clearly of the same
order of magnitude as the Cabibbo angle! 
Given the many uncertainties in this type of theory, the 
consistency of these results with Nature is remarkable. We note that the numerical 
value of the expansion parameter clearly depends on the contribution of 
the vector-like matter to $C_{\rm weak}$, about which we have no 
direct experimental information. 

In addition, the values of the 
mixed gravitational anomaly is also determined through the relation

\be  C_g=12{{C_{\rm weak}}\over {k_{\rm weak}}}\ .\ee
For integer $k_{\rm weak}$ and ${\bf n}=3$, this implies that $C_{\rm 
grav}=-216,-108,-72,\dots$ for $k_{\rm weak}=1,2,3\cdots$, 
to be compared  with the {\it visible} matter 
contribution to  $C_{\rm grav}=-80$.  Thus   additional 
fields are required, and  $k_{\rm weak}\le 2$, to avoid fields 
with positive $X$-charges that 
spoil the DSW vacuum. Another argument for  new fields is 
that not all anomalies are cancelled, since we have from the $\theta$ 
sector 

\be XXY^{(2)}=1 \ ;\qquad Y^{(1)}Y^{(1)}Y^{(2)}= Y^{(1)}Y^{(2)}Y^{(2)}=-1 
\ ,\ee
and from all {\it visible} matter
\be\qquad XY^{(1)}Y^{(2)}=-18 \ .\ee
The construction of a hidden sector theory that cancels these anomalies, 
and provides the requisite $C_{\rm grav}$ is rather arbitrary, since we  
have few guidelines: anomaly cancellation, and the absence of flat directions 
which indicates that the $X$ charges of the hidden matter should be 
negative.

If we use as a theoretical guide the $E_8\times E_8$ heterotic theory, we 
expect an exceptional gauge theory in the hidden sector. 
In particular, Bin\'etruy and Dudas~\cite{BD} considered a hidden gauge 
group G with a pair of matter fields with the same $X$-charge, but vector-like
with respect to all other symmetries, causing supersymmetry breaking. 
This theory contributes to few anomalies, only in  $C_{\rm grav}$, $(XY^{(1)}Y^{(2)})$ 
and the anomaly associated with the hidden
gauge group G, related by the Green-Schwarz relation

\be C_G=-18{k_G\over k_{\rm weak}}\ee 
where $k_G$ is the Kac-Moody integer level ($k_G$ integer heavily 
constrains  possible theories of this type). It must be augmented by other 
fields, since it does not cancel the remaining 
anomalies $(XXY^{(2)})$, $(Y^{(1)}Y^{(1)}Y^{(2)})$ and  
$(Y^{(1)}Y^{(2)}Y^{(2)})$.  These will be accounted for by singlet 
fields.

There is a simple set of four singlet fields, $\Sigma_a$ which absorb 
many of the remaining anomalies, without creating unwanted flat directions. 
Their charges are given in the following table:
\vskip 0.3cm

\hskip 2cm
\begin{center}
\begin{tabular}{|c|c|c|c|c|}

           \hline          
$ $&${\rm {\bf \Sigma _1}}$ & ${\bf \Sigma _2}$ & ${\bf \Sigma _3}$ & 
${\bf \Sigma _4}$  \\
 \hline \hline   

$X$& $-1/2$& $-1/2$ & ${0}$ &${0}$ \\

$Y^{(1)}$& $0$& $0$ & $1/2$&  $-1/2$ \\                                                            
 
$Y^{(2)}$&$-9/4$& $-7/4$ &$9/4$&$7/4$\\ \hline
 \end{tabular}
\end{center}
\vskip 0.3cm 
\noindent They cancel the anomalies from the $\theta$ sector, since
over the $\Sigma$ fields 
\be XY^{(1)}Y^{(2)}=0\ ,\ \  XXY^{(2)}=-1 \ ,\ \  Y^{(1)}Y^{(1)}Y^{(2)}=
Y^{(1)}Y^{(2)}Y^{(2)}=1\ ,\ee
as well as $ C_{\rm grav}=-1$.

The remaining anomalies can be accounted for by a simple gauge theory 
based on $G=E_6$; it has two matter fields with 
$X=x_1$, transforming as the 
$\bf 78$, and one $\bf 27$, $\bf{\overline {27}}$ pair, each with $X=x_2$. For $k_{\rm 
weak}=1$, we find that
\be C_{\rm grav}=-216~~~\rightarrow ~~~2(27x_2+78x_1)=-135\ .\ee
The gauge anomaly condition is given by
\be
C_G=2(6x_2+24x_1)=-18k_G\ .\ee
For $k_G=1$, one of the charge is positive, leading to undesirable flat 
directions, while for $k_G=2$, we find $x_1=-9/20$ and $x_2=-6/5$. 
The adjoint fields have no $Y^{(1,2)}$ charges, and the pair of $\bf 
27$-$\bf{\overline {27}}$ have vector-like with respect to $Y^{(1,2)}$, 
with charges $5/9$ and $1/2$, respectively. This sector breaks 
supersymmetry, but it cannot be the main agent for supersymmetry 
breaking, since it produces non-degenerate squark masses, and our model 
does not have alignment.

The singlet fields have little effect on low energy  phenomenology. 
Computation of the powers of the $\theta $ fields  in the mass 
invariants $\Sigma_a\Sigma_b$, yield in the DSW vacuum 
 the mass matrix of the $\Sigma$ fields before SUSY breaking 
\be M_{\Sigma}=\pmatrix {0&0&0&0\cr
                          0&0&0&0\cr
                          0&0&0&M\lambda _c^8\cr
                          0&0&M\lambda _c^8&0\cr}\ ;\ee
it has two zero eigenvalues. The Giudice-Masiero mechanism can fill in the 12 
(and 21) entries after SUSY breaking, yielding:

\be M_{\Sigma}=\pmatrix {0&m\lambda _c^7&0&0\cr
                          m\lambda _c^7&0&0&0\cr
                          0&0&0&M\lambda _c^8\cr
                          0&0&M\lambda _c^8&0\cr}\ ,\ee
where $m$ is of order of the SUSY breaking scale. 
The above matrix has now two large ($\sim 10^{11}\; GeV$) and two small 
($1-100 \; MeV$) eigenvalues. 
The two heavy states get diluted during inflation.
The two light states are stable since their lowest order coupling to the 
light fields is quartic, dominated by terms like $ \Sigma _1\Sigma _2 
H_uH_d$ . Although stable, and undiluted 
by inflation, their contribution to the  energy density of the universe 
is negligible. 

Finally we note that it is difficult to produce models for the hidden sector; 
for example  
we could take  $G=E_7$ with $k_G=2$, two matter fields 
transforming as the ${\bf 133}$ (adjoint) representation, 
but there does not seem to be any simple set of singlet fields 
with the requisite anomalies. 


\mysection{R-Parity}

The invariants of the minimal standard model and their associated flat 
directions have been analyzed in detail in the literature~\cite{GKM}. In 
models with an anomalous $U(1)$, these invariants carry in general  
$X$-charges, which,  as we have seen,  determines their suppression in 
the effective Lagrangian. Just as there is a basis of invariants,  proven 
long ago by Hilbert, the charges of these invariants are not all 
independent; they can in fact be expressed in terms of the charges of the 
lowest order invariants built out of the fields of the minimal standard 
model, and some anomaly coefficients. 

The $X$-charges of the three types of cubic standard model invariants that violate 
$R$-parity as well as baryon and/or lepton numbers can be expressed in 
terms of the $X$-charges of the MSSM invariants and the R-parity 
violating invariant
 
\be
  X^{[\barre R]}\equiv X(LH_u)\ ,\ee
through the relations

\be
X_{L{\bf Q}\bar{\bf d}}=X^{[d]}-X^{[\mu]}+X^{[\barre R]}\ 
,\nonumber\ee
 \be
X_{LL\bar e}=X^{[e]}-X^{[\mu]}+X^{[\barre R]}\ .\nonumber\ee
\be X_{\bar{\bf u}\bar{\bf d}\bar{\bf d}}=X^{[d]}+X^{[\barre R]}
+{1\over 3}\left(C_{\rm color}-C_{\rm weak}\right)-{2\over 3}X^{[\mu]}\ .\nonumber\ee
Although they vanish in our model, we still display 
 $X^{[u]}$ and $X^{[\mu]}=0$, since these sum rules are more general. 

In the analysis of the flat directions, we have seen how 
the seesaw mechanism forces the $X$-charge of 
$\overline N$ to be half-odd integer. Also, the Froggatt-Nielsen~\cite{FN} 
suppression of the minimal standard model invariants, and the holomorphy 
of the superpotential  require 
$X^{[u,d,e]}$ to be zero or negative integers, and the equality of the 
K\'ac-Moody levels of $SU(2)$ and $SU(3)$  forces $C_{\rm color}=C_{\rm 
weak}$, through the Green-Schwarz mechanism. Thus we conclude that the 
$X$-charges of these operators are half-odd integers, and thus they cannot 
appear in the superpotential unless multiplied by at least one 
$\overline N$. This reasoning can be applied to the  higher-order 
$\barre R$ operators since their charges are given by 

\bea
X_{{\bf Q}{\bf Q}{\bf Q}H_d}&=&
X^{[u]}+X^{[d]}-{1\over 3}X^{[\mu]}-X^{[\barre R]}\ ,\\ 
X_{\bar{\bf d}\bar{\bf d}\bar{\bf d}LL}&=&
2X^{[d]}-X^{[u]}-{5\over 3}X^{[\mu]}+3X^{[\barre R]}\ ,\\
X_{{\bf Q}{\bf Q}{\bf Q}{\bf Q}\bar{\bf u}}&=&
2X^{[u]}+X^{[d]}-{4\over 3}X^{[\mu]}-X^{[\barre R]}\ ,\\ 
X_{\bar{\bf u}\bar{\bf u}\bar{\bf u}\bar e\bar e}&=&
2X^{[u]}-X^{[d]}+2X^{[e]}-{2\over 3}X^{[\mu]}-X^{[\barre R]}\ ,
\eea
It follows that {\bf there are no $R$-parity violating operators, 
whatever their dimensions} : through the right-handed 
neutrinos, $R$-parity is linked to half-odd integer charges, so that 
charge invariance results in $R$-parity invariance. Thus  {\bf none} of 
the operators that violate $R$-parity can appear in holomorphic invariants: 
even after breaking of the anomalous $X$ symmetry, the remaining 
interactions all respect $R$-parity, leading to an {\bf absolutely 
stable superpartner}.  This is a general result deduced from  
the uniqueness of the DSW vacuum, the 
Green-Schwarz anomaly cancellations, and the seesaw mechanisms.


\mysection{Proton Decay}

In the presence of the extra discrete symmetry we introduced before, the 
operators that mix MSSM fields and vector-like matter and trigger proton 
decay
are excluded. Since R-parity is exactly conserved, the dangerous
dimension 3 operators $L{\bf Q}{\bf \overline d}$ and ${\bf \overline u}{\bf
\overline d}{\bf \overline d}$ that usually induce fast proton decay are also
excluded.
This leaves for the dominant sources of proton decay the dimension 5
operators that appear in the effective Lagrangian as 
\be W={1\over M}[{\kappa _{112i}}{\bf Q}_1{\bf Q}_1{\bf Q}_2{\bf
L}_i+{\overline \kappa} _{1jkl}
{\bf {\overline u}}_i{\bf {\overline u}}_j{\bf {\overline d}}_k{\bf {\overline 
e}}_l]\ee
where for the first operator the flavor index $i=1,2$ if there is a charged 
lepton in
the final state and $i=1,2,3$ if there is a neutrino and $j=2,3$,
$k,l=1,2$. We have denoted the suppression factors in the DSW 
vacum in front of the operators by $\kappa $ and ${\overline
\kappa}$. These operators could for example give rise  
the proton decay modes
$p\rightarrow \pi^{+}{\overline {\nu}}_{i}$ 
and 
$p\rightarrow \pi^{0}l^{+}_{i}$
or to $p\rightarrow K^{+}{\overline {\nu}}_{i}$ 
and 
$p\rightarrow K^{0}l^{+}_{i}$.   
In \cite{BNir}, the phenomenological limits on these
suppression factors were computed to be:
\be \kappa _{112i}\le \lambda _c^{11}\ee  
and
\be {\overline \kappa _{1jkl}}{(K^{u}_{RR})}_{1j}\le \lambda _c^{12}\ee
where ${K^{u}_{RR}}_{1j}=V^u_R{\tilde V}^{\dagger}_R$. $V_R$ are the 
matrices
that diagonalize on the right the quark and the squark matrices
respectively. We can easily calculate it in this model:
\be {K^{u}_{RR}}_{1j}=\pmatrix {1&\lambda _c^3&\lambda _c^5 \cr
                                \lambda _c^3&1&\lambda _c^2 \cr
                                \lambda _c^5&\lambda _c^2&1 \cr}.\ee
In table 3 we give in the first column a list of the dangerous operators 
${\bf 
Q}{\bf
Q}{\bf Q}L$ (${\bf {\overline u}}{\bf {\overline u}}{\bf {\overline d}}{\bf 
{\overline
e}}$) and in the second column the suppression $\kappa _{ijkl}$ (${\overline
\kappa _{ijkl}}K^{u}_{RR}$) that we computed in our model. 

\begin{table}
\vskip -2cm
\caption{{\bf Operators inducing proton decay and their suppression.}}
\vskip 1cm
\hskip 2cm
\begin{center}
\begin{tabular}{|c|c|}

           \hline  &    \\
$~{\rm {\bf Operator}}~$ & $ {\bf Supression} $\\ 
                       \hline \hline   &    \\

${\bf Q}_1{\bf Q}_1{\bf Q}_2{\bf L}_1$& 
$\lambda _c^{14}$ \\
                                      &    \\ 
                                                                 
${\bf Q}_1{\bf Q}_1{\bf Q}_2{\bf L}_{2,3}$& 
$\lambda _c^{11}$  \\ 
                                      &    \\
${\overline {\bf u}}_1{\overline {\bf u}}_2{\overline {\bf d}}_1{\overline {\bf 
e}}_{1}$& 
$\lambda _c^{15}$  \\ 
                                      &    \\

${\overline {\bf u}}_1{\overline {\bf u}}_2{\overline {\bf d}}_1{\overline {\bf 
e}}_{2}$& 
$\lambda _c^{16}$  \\ 
                                      &    \\

${\overline {\bf u}}_1{\overline {\bf u}}_2{\overline {\bf d}}_2{\overline {\bf 
e}}_{1}$& 
$\lambda _c^{14}$  \\ 
                                      &    \\

${\overline {\bf u}}_1{\overline {\bf u}}_2{\overline {\bf d}}_2{\overline {\bf 
e}}_{2}$& 
$\lambda _c^{15}$  \\ 
                                      &    \\

${\overline {\bf u}}_1{\overline {\bf u}}_3{\overline {\bf d}}_1{\overline {\bf 
e}}_{1}$& 
$\lambda _c^{13}$  \\ 
                                      &    \\

${\overline {\bf u}}_1{\overline {\bf u}}_3{\overline {\bf d}}_1{\overline {\bf 
e}}_{2}$& 
$\lambda _c^{14}$  \\ 
                                      &    \\

${\overline {\bf u}}_1{\overline {\bf u}}_3{\overline {\bf d}}_2{\overline {\bf 
e}}_{1}$& 
$\lambda _c^{12}$  \\ 
                                      &    \\

${\overline {\bf u}}_1{\overline {\bf u}}_3{\overline {\bf d}}_2{\overline {\bf 
e}}_{2}$& 
$\lambda _c^{13}$  \\ 
                                      &    \\

\hline

\end{tabular}
\end{center}
\end{table}
%

Even though all operators in table 3 seem naively sufficiently
suppressed so that proton decay is within the experimental bound,
it is interesting to examine them more closely from the phenomenological 
point of view. Consider the operator ${\bf Q}_1{\bf Q}_1{\bf
Q}_2{\bf L}_2$. This operator can lead to proton decay via a wino,
gluino, zino, photino or Higgsino 
exchange. The contribution
via gluino exchange could be the dominant due to the strong coupling of the
gluino. Here let us recall that experimental data strongly suggests a near
degeneracy between squark masses in order to avoid large contributions
to Flavor Changing Neutral Currents (FCNC). One mechanism that has
been suggested~\cite{alignment} is where alignement between quarks
and squarks takes place and therefore FCNC are suppressed
irrespectively of the SUSY breaking mechanism. One can calculate
in the model the extent of such an alignment. We find that there is
no sufficient quark-squark alignement and therefore  FCNC are
not sufficiently suppressed. To agree with experimental
data we have to assume that the squark masses that result from SUSY
breaking are approximately degenerate, a fact that does not seem
to be unlikely in the context of realistic superstring models
\cite{degeneracy}. 
In such a case, the contribution due to gluino exchange is negligible. 
\par Generically, a careful calculation of a proton decay process not 
only involves
uncertainties due to our ignorance of superpartner masses but also due to 
large uncertainties 
in hadronic matrix elements. Assuming nearly degenerate squarks, 
 the dominant decay mode is via wino exchange
and  the decay rate for the process $p\rightarrow K^0\mu ^+$ is 
given by \cite{Protondecay}:
\begin{equation} \Gamma (p\rightarrow K^0\mu ^+)=(  {{10.5}{b} \alpha_2 
\cos 
{\theta _c}\over {\pi M_{}}})^2
{(m_p^2-m_K^2)^2\over {8\pi m_p^3f_{\pi}^2}}|{0.7\kappa 
_{1122}
{f(m_{\tilde w},m_{\tilde q}})}|^2 \end{equation} 
were here $b =(0.003-0.03)~GeV^3$ is an unknown strong matrix element, 
$\alpha _2=\alpha /\sin ^2{\theta _W}$ and from our earlier
estimates of the cut-off, $M\sim3 \times 10^{16}~GeV$.
We have two regimes to consider 
\be \;\;m_{\tilde w} << m_{\tilde q}:\;\;\;f(m_{\tilde w},m_{\tilde 
q})={m_{\tilde w}\over m^2_{\tilde q}}\ ,\ee
and 
\be m_{\tilde w} >>m_{\tilde q}:\;\;\;f(m_{\tilde w},m_{\tilde 
q})={1\over m_{\tilde w}}\ln{m^2_{\tilde w}\over m^2_{\tilde
q}}\ .\ee
The experimental bound on the decay $p~\rightarrow~K^0+\mu^+$, which
is the dominant one in our theory, is~\cite{PDG}
\be \Gamma (p\rightarrow K^0 \mu ^+)< 
10^{32}\;{\rm years}^{-1}\ .\ee 
For wino mass much larger than squark masses, this decay rate is
several orders of magnitude lower than the experimental limit. For
wino masses much lower than squark masses, the rate is near the experimental
limit. For example, with $m_{\tilde w}\sim 100$ GeV, $m_{\tilde q}\sim 
800 $GeV, and $b=.003$, we get the  lifetime  $\sim  10^{31}$ years, 
near  the experimental bound. Unfortunately our model cannot be more 
precise, because of the unknown prefactors of order one terms in 
the effective interactions; Still it predicts that the proton decays preferentially 
into a neutral
$K$ and an antimuon with a lifetime at or near the present experimental limit. 
 Finally we note that if we use the expansion parameter determined 
through the Green-Schwarz relation, and not the Cabibbo angle, 
our estimates get worse and our model implies a proton lifetime slightly 
shorter than the experimental bound. As we remarked earlier, this value 
of the expansion parameter depends on the contribution of the vector-like matter 
to $C_{\rm weak}$.


\mysection{Flat direction analysis}

Our model is now completely specified, except for the supersymmetry breaking
sector. We can study its flat directions and check whether the DSW
vacuum is unique, using the techniques introduced in Ref. \cite{IL}. We
shall only sketch the  main points, and refer the interested reader to this
reference for more details and the discussion of some subtleties.

In the presence of an anomalous $U(1)$, the well-known correspondence
between the zeroes of the $D$-terms and the holomorphic gauge invariants
\cite{Buccella} breaks down. However, the existence of the DSW vacuum
$\svev{\theta_1} = \svev{\theta_2} = \svev{\theta_3} = \xi^2$ allows us
to rewrite the Abelian $D$-term constraints as:

\begin{equation}
  \left( \begin{array}{c} \svev{\theta_1} - \xi^2 \\ \svev{\theta_2} -\xi^2 \\
\svev{\theta_3} - \xi^2 \end{array} \right)\ =\ \sum_a\ v^2_a\ \left(
\begin{array}{c} n^a_1 \\ n^a_2 \\ n^a_3 \end{array} \right)\ +\ \sum_i\
\svev{\chi_i}\ \left( \begin{array}{c} n^i_1 \\ n^i_2 \\ n^i_3 \end{array}
\right)\ ,
\label{eq:Dterms}
\end{equation}
where the \{$\chi_i$\} are standard model singlets other than the $\theta$
fields, the $v^2_a$ are vevs associated with a basis of standard model
invariants \{$S_a$\}, and the numbers $n^a_{\alpha}$ (resp. $n^i_{\alpha}$)
are associated with the invariant $S_a$ (resp. singlet $\chi_i$) by Eq.
(\ref{eq:powers}). In the present model, the $\chi$ fields are the three
right-handed neutrinos $\overline{N}_1$, $\overline{N}_2$, $\overline{N}_3$
and the $\Sigma$ fields needed to ensure anomaly
cancellation\footnote{As eluded to earlier, we have not included the
$SO(10)$ singlets $S_1$, $S_2$, $S_3$ necessary to make up three complete
families in the ${\bf 27}$ of $E_6$; otherwise the superpotential would
contain an invariant $S_1\, \theta^3_2 \theta^3_3$ linear in $S_1$, which
would spoil the DSW vacuum.}.
The basis of standard model invariants includes the MSSM
basis of Ref. \cite{GKM} as well as invariants containing the vector-like
fields, such as the ones discussed in section \ref{section:VL}. Eq.
(\ref{eq:Dterms}) tells us that $D$-flat directions are parametrized by
the vacuum expectation values of both the standard model invariants and
the $\chi$ fields. The generic effect of $F$-term contraints and
supersymmetry breaking is to fix these vevs, resulting in a particular
low-energy vacuum. As stressed in Ref. \cite{IL}, the computation of the
$n_{\alpha}$ simplifies a lot the discussion of $D$- and $F$-flatness.

Consider first the flat directions involving only standard model singlets.
Assuming for simplicity that only one $\chi$ field acquires a vev, we must
distinguish between two cases:
\begin{itemize}
  \item all $n_{\alpha}$ are positive. Then $\svev{\theta_\alpha} \geq
\xi^2$ for $\alpha=1,2,3\,$, whatever $\vev{\chi}$ may be. In addition,
the superpotential contains an invariant of the form $\chi^m\, \theta^{m_1}_1
\theta^{m_2}_2 \theta^{m_3}_3$, with $m_\alpha = m\, n_\alpha$ (as discussed
in Section \ref{section:FD}, $m \geq 2$ is required in order not to spoil the
DSW vacuum). The $F$-term constraints then impose $\vev \chi = 0\,$: the
flat direction is lifted down to the DSW vacuum.
  \item some of the $n_\alpha$ are negative. The relations
$\svev{\theta_\alpha} \geq \xi^2$ no longer hold, and the low-energy
vacuum may be different from the DSW vacuum. In our model, this happens
only for $\overline N_3$, for which ($n_1$, $n_2$, $n_3$) =  ($3/2$,
$-1/2$, $1/2$). One can then see from (\ref{eq:Dterms}) that the vacuum
$\vev{\overline N_3, \theta_1, \theta_3}$ with $\svev{\overline N_3}
= 2\, \xi^2$, $\svev{\theta_1} = 4\, \xi^2$ and
$\svev{\theta_3} = 2\, \xi^2$
is perfectly allowed by $D$-term constraints. This is a rather unwelcome
feature, because most Yukawa couplings vanish in this vacuum. Fortunately,
the superpotential contains an invariant $\overline N^3_3\, \overline N_2\,
\theta^6_1 \theta^4_3$, with no power of $\theta_2$, which lifts the
undesired vacuum.
\end{itemize}

This discussion can be generalized to flat directions involving several
$\chi$ fields; we conclude that the model does not possess any other stable
vacuum of singlets than the DSW vacuum. Thus, the low-energy mass
hierarchies are completely determined by the symmetries at high energy.

Flat directions involving fields charged under $SU(3)_C \times SU(2)_L
\times U(1)_Y$ can be analyzed in a similar way. For each element $S$ of
the basis of invariants, we compute the numbers ($n_1$, $n_2$, $n_3$). If
one of the $n_\alpha$ is negative, we must check that the superpotential
contains a term of the form $S'\, \theta^{n'_1}_1 \theta^{n'_2}_2
\theta^{n'_3}_3$ (with $S'$ a combination of basis $G$-invariants and
$\chi$ fields), where either one of the following two conditions is
fulfilled: (i) $S'$ contains no other field than the ones appearing in $S$,
and $n'_{\alpha}=0$ or $1$, if $n_{\alpha}<0$ (with the additional
constraint that no more than one such $n'_{\alpha}$ should be equal to $1$);
(ii) $S'$ contains only one field that does not appear in $S$, and
$n'_{\alpha}=0$, if $n_{\alpha}<0$. This ensures that there is no flat
direction associated with the single invariant $S$.

Remarkably enough, those conditions are always fulfilled in our model,
despite the great number of standard model invariants. In Table
\ref{caption:FD_MSSM}, we list the MSSM basis invariants for which some
of the $n_\alpha$ are negative. For each of these invariants (first column),
we give the corresponding numbers $n_1$, $n_2$, $n_3$ (second column),
the associated flat direction that breaks the standard model symmetries
(third column), and an invariant that lifts it (fourth column).

The case of flat directions involving vector-like matter is slightly
different. Since we have assumed the existence of a discrete symmetry
that prevents numerous invariants from appearing in the superpotential,
there could be flat directions associated with these invariants. But this
is not the case, as long as the vector-like fields are massive. Their
$F$-terms take indeed the following form (gauge indices are not shown,
and powers of the $\theta$ fields have been absorbed in the mass matrices
for simplicity):

\begin{equation}
  F_{\overline{E}_i}\ =\ M_{\overline{E}_i E_j}\, E_j\ +\ \ldots \qquad
  F_{\overline{\bf D}_i}\ =\ M_{\overline{\bf D}_i \bf D_j}\, \bf D_j\ +\ \ldots
\label{eq:F_E}
\end{equation}
\begin{equation}  \hskip -.9cm
F_{E_j}\ =\ M_{\overline{E}_i E_j}\, \overline{E}_i\ +\ \ldots \qquad
  F_{\bf D_j}\ =\ M_{\overline{\bf D}_i \bf D_j}\, \overline{\bf D}_i\ +\ \ldots
\label{eq:F_Ebar}
\end{equation}
where the dots stand for possible higher order contributions. Since the
matrices $M_{\overline{E} E}$ and $M_{\overline{\bf D} \bf D}$ are invertible,
one concludes that the vanishing of (\ref{eq:F_E}) and (\ref{eq:F_Ebar})
forbids any flat direction involving vector-like fields, provided that
it is associated with an invariant for which all $n_\alpha$ are positive.
That this is true also for invariants with one or several negative
$n_\alpha$ is less obvious. It is due to the following features of the
model: the (1,1) entry of the vector-like mass matrices are generated from
the superpotential terms $\overline{E}_1 E_1\, \theta^3_1$ and
$\overline{\bf D}_1 \bf D_1\, \theta^3_1$, and all invariants that have one
or several negative $n_\alpha$ both satisfy $n_1 \geq 0$ and contain
at least one vector-like field of the first family. Therefore, condition
(ii) is always fulfilled. This can be checked in Table \ref{caption:FD_VL}
(where only operators up to superfield dimension 4 have been displayed).

We have thus
checked that the superpotential contains terms that lift all flat
directions associated with a single standard model invariant. This is not
sufficient, however, to ensure that the standard model symmetries are not
broken at the scale $\xi$. Other invariants than those of Tables
\ref{caption:FD_MSSM} and \ref{caption:FD_VL} are in general necessary
to lift completely the flat directions associated with several standard
model invariants and singlets. While we did not perform a complete
analysis - which would be rather tedious -, it is clear that most, if 
not all, flat directions are forbidden by the $F$-term constraints.

We conclude that the vacuum structure of our model is satisfactory: 
the only stable vacuum of singlets
allowed by $D$- and $F$-term constraints is the DSW vacuum, and 
flat directions associated with a single $SU(3)_C \times SU(2)_L
\times U(1)_Y$ invariant are lifted by the $F$-terms.
The only expected effects of supersymmetry breaking are to lift the
possible remaining flat directions, and to shift slightly the DSW vacuum
by giving a small or intermediate vev to other singlets or to fields
with standard model quantum numbers.


\mysection{Conclusion}
We have presented a simple model that extends the standard model gauge 
group by three phase symmetries, one of which is anomalous. The extra 
symmetries are broken in the DSW vacuum, thereby providing a small 
{\it computable} expansion parameter, in terms of which the Yukawa couplings of the 
standard model can be expanded. The model has a natural cut-off 
characterized by the scale at which the anomalies are absorbed by the 
Green-Schwarz terms, which  is the gauge unification scale. The expansion 
parameter, which  depends on the contribution of the standard-model vector-like 
matter to the weak anomaly, turns out to be close to the Cabibbo angle. 
All Yukawa hierarchies as well as the Weinberg angle are reproduced if 
the expansion parameter is taken to be the Cabibbo angle. The model is 
predictive in the neutrino sector, yielding three massive neutrinos with 
small mixings between the electron neutrino and the muon and tau 
neutrinos, and mixings of order one between the muon and tau neutrinos. 
With the cut-off near the unification scale, the solar neutrino deficit 
is explained in terms of the non-adiabatic MSW effect, and the 
atmospheric neutrino imbalance is reproduced. With the Cabibbo angle as 
expansion parameter, our model is compatible with proton decay bounds.

Many of the uncertainties of the model are associated with the nature of 
its vector-like matter, which determines gauge unification and the value of the 
expansion parameter. In addition, it must contain  matter with no 
standard-model charges, to cancel anomalies. Although we made a definite 
proposal for those fields, our lack of experimental guidelines should 
be kept in mind. Our model shows the way in which many of the generic features 
encountered in the compactification of theories in higher dimensions 
can be used to deduce phenomenological constraints. 
Finally we note that the value of the cut-off is the gauge unification, 
underlining the well-known possible conflict with compactified string 
theories.

\vskip 1cm 
{\bf Acknowledgements}
\vskip .5cm

We acknowledge usefull discussions with S. Chang. S.L. thanks the Institute
for Fundamental Theory, Gainesville, for its hospitality and financial
support.




\begin{table}
\vskip -2cm
\caption{{Flat Directions of MSSM and ${\overline N}$ fields.}}
\label{caption:FD_MSSM}
\vskip 1cm
\hskip -0.5cm
\begin{center}
\begin{tabular}{|c|c|c|c|}

           \hline  & & &  \\
$~{\rm {\bf Basis ~Invariant}}~$ & $ {\bf (n_1,n_2,n_3)} $ & $~{\bf Flat~ 
Direction}~$ 
& 
$~{\bf ~FD~ lifted~ by}~$\\  & & &  \\
                       \hline \hline   & & &  \\
$~{\overline N}_3$&$ (3/2,-1/2,1/2) ~$ & 
$~<{\overline N}_3,\theta_1,\theta_3>~$ & 
${\overline N}_2{\overline N}_3^{\, 3}\, \theta_1^{6}\theta_3^{4}~$ \\
                                      & & &  \\ 
                                                                 
${\bf L}_1{\bf H_u}$&$ (-3/2,1/2,5/2) ~$ & 
$~<{\bf L}_1,{\bf H_u},\theta_2,\theta_3>~$ & 
$~{\bf L}_1{\overline N}_3{\bf H_u}\theta_3^{3}~$\\ 
                                      & & &  \\
${\bf L}_2{\bf H_u}$&$ (-3/2,1/2,-1/2)~$ & 
$~<{\bf L}_2,{\bf H_u},\theta_2,\theta_3>~$ & 
$~{\bf L}_2{\overline N}_3{\bf H_u}~$\\           
                                                             & & &  \\  
${\bf L}_3{\bf H_u}$&$ (-3/2,1/2,-1/2)~$ & 
$~<{\bf L}_3,{\bf H_u},\theta_2,\theta_3>~$ & 
$~{\bf L}_3{\overline N}_3{\bf H_u}~$\\ 

                                                        & & &  \\  

${\bf L}_2{\overline 
e}_1{\bf H_d}$&$ (3,2,-1)~$ &
$<{\bf L}_2,{\overline e}_1,{\bf H_d},\theta_1,\theta_2>~$ &
${\bf L}_2{\overline e}_3{\bf H_d}\theta_1^{3}$\\
 
                                                                 & & &  \\

${\bf L}_3{\overline 
e}_1{\bf H_d}$&$ (3,2,-1)~$ &
$<{\bf L}_3,{\overline e}_1,{\bf H_d},\theta_1,\theta_2>~$ &
${\bf L}_3{\overline e}_3{\bf H_d}\theta_1^{3}$\\
 
                                                                 & & &  \\

${\bf L}_2{\bf L}_3{\overline 
e}_1$&$ (3/2,5/2,-3/2)~$ &
$<{\bf L}_2,{\bf L}_3,{\overline e}_1,\theta_1,\theta_2>~$ &
${\bf L}_2{\bf L}_3{\overline e}_1{\overline 
N}_1\theta_1^{3}\theta_2^{6}$\\
 
                                                                 & & &  \\
${\bf L}_2{\bf L}_3{\overline 
e}_3$&$ (3/2,1/2,-1/2)~$ &
$<{\bf L}_2,{\bf L}_3,{\overline e}_3,\theta_1,\theta_2>~$ &
${\bf L}_2{\bf L}_3{\overline e}_3{\overline N}_3\theta_1^{3}$\\  
 
                                                     & & &  \\

${\bf L}_2{\bf Q}_3{\bf {\overline 
d}_2}$&$ (3/2,1/2,-1/2)~$ &
$<{\bf L}_2,{\bf Q}_3,{\bf {\overline d}_2},\theta_1,\theta_2>~$ &
${\bf L}_2{\bf Q}_3{\bf {\overline d}_2}{\overline N}_3
\theta_1^{3}$\\ 
                                                                & & &  \\  
${\bf L}_3{\bf Q}_3{\bf {\overline 
d}_2}$&$ (3/2,1/2,-1/2)~$ &
$<{\bf L}_3,{\bf Q}_3,{\bf {\overline d}_2},\theta_1,\theta_2>~$ &
${\bf L}_3{\bf Q}_3{\bf {\overline d}_2}{\overline N}_3
\theta_1^{3}$\\ 
 
                                                                 & & &  \\  
${\bf L}_2{\bf Q}_3{\bf {\overline 
d}}_3$&$ (3/2,1/2,-1/2)~$ &
$<{\bf L}_2,{\bf Q}_3,{\bf {\overline d}}_3,\theta_1,\theta_2>~$ &
${\bf L}_2{\bf Q}_3{\bf {\overline d}}_3{\overline N}_3\theta_1^{3}$\\ 
                                                                & & &  \\
${\bf L}_3{\bf Q}_3{\bf {\overline 
d}}_3$&$ (3/2,1/2,-1/2)~$ &
$<{\bf L}_3,{\bf Q}_3,{\bf {\overline d}}_3,\theta_1,\theta_2>~$ &
${\bf L}_3{\bf Q}_3{\bf {\overline d}}_3{\overline N}_3\theta_1^{3}$\\  
                                                                & & &  \\
${\bf {\overline u}}_3{\bf {\overline d}}_2
{\bf {\overline d}}_3$&$ (3/2,1/2,-1/2)~$ &
$<{\bf {\overline u}}_3,{\bf {\overline d}}_2,
{\bf {\overline d}}_3,\theta_1,\theta_2>~$ 
&
${\bf {\overline u}}_3{\bf {\overline d}}_2{\bf {\overline d}}_3
{\overline N}_3\theta_1^{3}$\\ 
                                                                & & &  \\  
   
${\bf Q}_3{\bf {\overline u}}_3{\overline e}_1{\bf 
H_d}$&$ (9/2,3/2,-1/2)~$ &
$<{\bf Q}_3,{\bf {\overline u}}_3,{\overline e}_1,{\bf 
H_d},\theta_1,\theta_2>~$ 
&
${\bf Q}_3{\bf {\overline u}}_3{\bf H_u}$\\ 
                                                                & & &  \\  
${\bf Q}_3{\bf {\overline u}}_3{\overline e}_3{\bf 
H_d}$&$ (9/2,-1/2,1/2)~$ &
$<{\bf Q}_3,{\bf {\overline u}}_3,{\overline e}_3,{\bf 
H_d},\theta_1,\theta_3>~$ 
&
${\bf Q}_3{\bf {\overline u}}_3{\bf H_u}$\\
                                                                & & &  \\   
${\bf Q}_3{\bf {\overline u}}_3{\bf L}_2{\overline 
e}_1$&$ (3,2,-1)~$ &
$<{\bf Q}_3,{\bf {\overline u}}_3,{\bf L}_2,{\overline 
e}_1,\theta_1,\theta_2>~$ 
&
${\bf Q}_3{\bf {\overline u}}_3{\bf H_u}$\\
                                                                & & &  \\  
${\bf Q}_3{\bf {\overline u}}_3{\bf L}_3{\overline 
e}_1$&$ (3,2,-1)~$ &
$<{\bf Q}_3,{\bf {\overline u}}_3,{\bf L}_3,{\overline 
e}_1,\theta_1,\theta_2>~$ 
&
${\bf Q}_3{\bf {\overline u}}_3{\bf H_u}$\\
                                                                & & &  \\ 
${\bf Q}_3{\bf {\overline u}}_3{\bf Q}_3{\bf {\overline u}}_3{\overline 
e}_1$&$ (9/2,3/2,-1/2)~$ &
$<{\bf Q}_3,{\bf {\overline u}}_3,{\overline e}_1,\theta_1,\theta_2>~$ &
${\bf Q}_3{\bf {\overline u}}_3{\bf H_u}$\\
                                                                & & &  \\  
${\bf Q}_3{\bf {\overline u}}_3{\bf Q}_3{\bf {\overline u}}_3{\overline 
e}_3$&$ (9/2,-1/2,1/2)~$ &
$<{\bf Q}_3,{\bf {\overline u}}_3,{\overline e}_3,\theta_1,\theta_3>~$ &
${\bf Q}_3{\bf {\overline u}}_3{\bf H_u}$\\
                                                                 & & &  \\   
${\bf {\overline d}}_1{\bf {\overline d}}_2{\bf {\overline d}}_3{\bf 
L}_2{\bf 
L}_3
$&$ (3/2,3/2,-1/2)~$ &
$<{\bf {\overline d}}_1,{\bf {\overline d}}_2,{\bf {\overline d}}_3,
{\bf L}_2,{\bf L}_3,\theta_1,\theta_2>~$ &
${\bf {\overline d}}_1{\bf {\overline d}}_2{\bf {\overline d}}_3
{\bf L}_2{\bf L}_3{\overline N}_3 \theta_1^{3}\theta_2~$\\  
 
                                                      & & &  \\     \hline

\end{tabular}
\end{center}
\end{table}



\begin{table}
\vskip -2cm
\caption{{\bf Flat Directions involving vector-like matter 
in the discrete symmetry scenario (up to quartic operators).}}
\label{caption:FD_VL}
\vskip 1cm
\hskip -0.5cm
\begin{center}
\begin{tabular}{|c|c|c|c|}

           \hline  & & &  \\
$~{\rm {\bf Basis ~Invariant}}~$ & $ {\bf (n_1,n_2,n_3)} $ & $~{\bf Flat~ 
Direction}~$ 
& 
$~{\bf FD~lifted~by}~$\\  & & &  \\
                       \hline \hline   & & &  \\

${\overline E}_1H_d$&$ (3,-1,-1)$ & 
$<{\overline E}_1,H_d,\theta_1>$ & 
$~{\overline E}_1E_1\, \theta_1^{3}~$ \\
                                      & & &  \\ 

${\bf D}_1{\bf \bar d}_{1}$&$(3/2,-1/2,3/2)$ & 
$<{\bf D}_1,{\bf \bar d}_{1},\theta_1,\theta_3>$ & 
$~{\overline D}_1D_1\, \theta_1^{3}~$\\ 
                                      & & &  \\
                                   
${\bf D}_1{\bf \bar d}_{2,3}$&$(3/2,-1/2,1/2)$ & 
$<{\bf D}_1,{\bf \bar d}_{2,3},\theta_1,\theta_3>$ & 
$~{\overline D}_1D_1\, \theta_1^{3}~$\\ 
                                      & & &  \\

${\bf L}_1{\overline E}_1$&$(3/2,-1/2,3/2)$ & 
$<{\bf L}_1,{\overline E}_1,\theta_1,\theta_3>$ & 
$~{\overline E}_1E_1\, \theta_1^{3}~$\\           
                                                             & & &  \\  

${\bf L}_{2,3}{\overline E}_1$&$ (3/2,-1/2,-3/2)~$ & 
$<{\bf L}_{2,3},{\overline E}_1,\theta_1,\theta_2>$ & 
$~{\overline E}_1E_1\, \theta_1^{3}~$\\
                                                        & & &  \\  

${\bf Q}_3{\bf \overline D}_1H_d$&$ (9/2,1/2,-1/2)$ &
$<{\bf Q}_3,{\bf \overline D}_1,H_d,\theta_1,\theta_2>$ &
$~{\overline D}_1D_1\, \theta_1^{3}~$\\
                                                                 & & &  \\

${\bf Q}_3{\bf \bar u}_3{\overline E}_1$&$ (3,-1,-1)~$ &
$<{\bf Q}_3,{\bf \bar u}_3,{\overline E}_1,\theta_1>$ &
$~{\overline E}_1E_1\, \theta_1^{3}~$\\
                                                                 & & &  \\

${\bf L}_{2,3}{\bf Q}_3{\bf \overline D}_1$&$ (3,1,-1)$ &
$<{\bf L}_{2,3},{\bf Q}_3,{\overline D}_1,\theta_1,\theta_2>$ &
$~{\overline D}_1D_1\, \theta_1^{3}~$\\
                                                                 & & &  \\

${\bf D}_1{\bf \bar u}_3{\bar e}_3$&$ (3,-1,1)$ &
$<{\bf D}_1,{\bf \bar u}_3,{\bar e}_3,\theta_1,\theta_3>$ &
$~{\overline D}_1D_1\, \theta_1^{3}~$\\
                                                                & & &  \\

${\bf \bar u}_3{\bf \bar d}_{2,3}{\bf \overline D}_1$&$ (3,1,-1)$ &
$<{\bf \bar u}_3,{\bf \bar d}_{2,3},{\bf \overline D}_1,\theta_1,\theta_2>$ &
$~{\overline D}_1D_1\, \theta_1^{3}~$\\ 
                                                        & & &  \\
  
${\bf Q}_3{\bf \bar u}_3{\bf Q}_3{\bf \overline D}_1$&$ (9/2,1/2,-1/2)$ &
$<{\bf Q}_3,{\bf \bar u}_3,{\bf Q}_3,{\bf \overline D}_1,\theta_1,\theta_2>$ &
$~{\overline D}_1D_1\, \theta_1^{3}~$\\
                                                              & & &  \\  

${\bf \overline D}_1{\bf \bar u}_2{\bf \bar u}_3{\bar e}_1$&$ (9/2,7/2,-1/2)$ &
$<{\bf \overline D}_1,{\bf \bar u}_2,{\bf \bar u}_3,
{\bar e}_1,\theta_1,\theta_2>$ &
$~{\overline D}_1D_1\, \theta_1^{3}~$\\
                                                              & & &  \\ 
 
${\bf Q}_3{\bf D}_1{\bf D}_2{\bf \overline E}_1$&$ (9/2,-1/2,1/2)$ &
$<{\bf Q}_3,{\bf D}_1,{\bf D}_2,{\bf \overline E}_1,\theta_1,\theta_3>$ &
$~{\overline D}_1D_1\, \theta_1^{3}~$\\
                                                      & & &  \\     \hline

\end{tabular}
\end{center}
\end{table}



\begin{thebibliography}{Ref}

\bibitem{GS} M. Green and J. Schwarz, Phys. Lett. B149 (1984) 117.
\bibitem{DSW} M. Dine, N. Seiberg and E. Witten, Nucl. Phys. B289 (1987) 
589; J. Atick, L. Dixon and A. Sen, Nucl. Phys. B292 (1987) 109.
\bibitem{Ib} L. Ib\'a\~nez, Phys. Lett. B303 (1993) 55.
\bibitem{IR} L. Ib\'a\~nez and G. G. Ross, Phys. Lett. B332 (1994) 100.
\bibitem{BR} P. Bin\'etruy and P. Ramond, Phys. Lett. B350 (1995) 49;
P. Bin\'etruy, S. Lavignac,  and P. Ramond,  Nucl. Phys. B477 (1996) 353.
\bibitem{Nir} Y. Nir, Phys. Lett. B354 (1995) 107.
\bibitem{JS} V. Jain and R. Shrock, Phys. Lett. B352 (1995) 83.
\bibitem{PMR} P. Ramond, Kikkawa Proceedings, 1996.hep-ph/9604251
\bibitem{SEESAW} M. Gell-Mann, P. Ramond, and R. Slansky in Sanibel
Talk, CALT-68-709, Feb 1979, and in {\it Supergravity} (North Holland,
Amsterdam 1979). T. Yanagida, in {\it Proceedings of the Workshop on
Unified Theory and Baryon Number of the Universe}, KEK, Japan, 1979.
\bibitem{BLIR} P. Bin\'etruy, S. Lavignac,  Nikolaos Irges,  and P. Ramond, 
Phys. Lett. B403 (1997) 38.
\bibitem{EIR} John K. Elwood, Nikolaos Irges, and P. Ramond, Phys. Lett.
B413 (1997) 322.
\bibitem{BD} P. Binetruy, E. Dudas, Phys. Lett. B389 (1996), 503.
G. Dvali and A. Pomarol, .
\bibitem{LNS} M. Leurer, Y. Nir, and N. Seiberg, Nucl. Phys. B398
(1993) 319, Nucl. Phys. B420 (1994) 468.
\bibitem{faraggi} A. Faraggi, Nucl. Phys. B387 (1992) 239, {\em ibid.} B403 
(1993) 101, {\em ibid.} B407 (1993) 57.
\bibitem{degeneracy} A. E. Faraggi and J. C. Pati, UFIFT-HEP-97-29,
hep-ph/9712516.
\bibitem{ESIX} F. G\"ursey, P. Ramond and P. Sikivie, Phys. Lett. 60B
(1976) 177.
\bibitem{GM} G. Giudice and A. Masiero, Phys. Lett. B206 (1988) 480;
V.S. Kaplunovsky and J. Louis, Phys. Lett. B306 (1993) 269.
\bibitem{IL}  N. Irges and S. Lavignac, preprint UFIFT-HEP-97-34,
hep-ph/9712239, to be published in Phys. Letters.
\bibitem{MNS} Z. Maki, M. Nakagawa and S. Sakata, Prog. Theo. Phys. 28
(1962) 247.
\bibitem{MSW} S.P. Mikheyev and A.Yu. Smirnov,
Yad. Fiz. {\bf 42}, 1441 (1985) [Sov. J. Nucl. Phys. {\bf 42}, 913 (1985)];
Il Nuovo Cimento C {\bf 9}, 17 (1986); L. Wolfenstein,Phys. Rev. D {\bf 17}, 2369 (1978);
Phys. Rev. D {\bf 20}, 2634 (1979).
\bibitem{SuperK} E. Kearns, talk at the ITP conference on Solar Neutrinos:
News about SNUs, December 2-6, 1997.
\bibitem{Soudan} S.M. Kasahara et al., Phys. Rev. D55 (1997) 5282.
\bibitem{BLPR} P. Bin\'etruy, S. Lavignac, S. Petcov and P. Ramond, 
Nucl. Phys. B496 (1997) 3.
\bibitem{Langacker} N. Hata and P. Langacker, Phys. Rev. D56 (1997) 6107.
\bibitem{atmospheric} M.C. Gonzalez-Garcia, H. Nunokawa, O.L.G. Peres,
T. Stanev and J.W.F. Valle, preprint hep-ph/9801368. 
\bibitem{LSND} C. Athanassopoulos et al., Phys. Rev. Lett. 75 (1995) 2650;
Phys. Rev. Lett. 77 (1996) 3082; preprint nucl-ex/9706006.
\bibitem{Precision} P.H. Frampton, M. Harada, IFP-748A-UNC, hep-ph/9711448.
\bibitem{Quarksinglets} V. Barger, M.S. Berger, R.J.N. Phillips,
Phys.Rev.D52 (1995) 1663-1683. 
\bibitem{BNir} V. Ben-Hamo and Y. Nir, Phys. Lett. B339 (1994) 77.
\bibitem{Relics} A. Faraggi, Nucl. Phys. B477 (1996) 65. 
\bibitem{GKM} T. Gherghetta, C. Kolda and S.P. Martin, Nucl. Phys. B468 
(1996) 37.
\bibitem{FN} C.~Froggatt and H.~B.~Nielsen Nucl. Phys. B147 (1979) 277.
\bibitem{alignment} Y. Nir and N. Seiberg, Phys. Lett. B309, (1993),337.
\bibitem{Protondecay} H. Murayama, D.B. Kaplan, Phys.Lett.B336
(1994), 221-228.
\bibitem{Buccella} F. Buccella, J.-P. Derendinger, S. Ferrara
and C.A. Savoy, Phys. Lett. B115 (1982) 375.








\bibitem{PDG} Review of Particle Properties. Particle Data Group,
Phys. Rev. D54 (1996), 83.




\end{thebibliography}
\end{document}